\documentclass[5p,times,sort&compress]{elsarticle}

\usepackage[T1]{fontenc}
\usepackage[utf8]{inputenc}
\usepackage[english]{babel}
\usepackage{textcomp}
\usepackage{microtype}
\usepackage{graphicx}
\usepackage{float}
\usepackage{subfig}
\usepackage{amsmath}
\usepackage{amssymb}
\usepackage{amsthm}
\usepackage{booktabs}
\usepackage{multirow}
\usepackage{array}
\usepackage{tabularx}
\usepackage{colortbl}
\usepackage{xcolor}
\usepackage{xfrac}
\usepackage{siunitx}
\usepackage{algorithm}
\usepackage{algpseudocode}
\usepackage{listings}
\usepackage{enumitem}
\usepackage{verbatim}
\usepackage{url}
\usepackage[unicode=true,
  bookmarks=true,
  bookmarksnumbered=false,
  bookmarksopen=false,
  breaklinks=true,
  pdfborder={0 0 1},
  backref=false,
  colorlinks=true,
  linkcolor=darkred,
  citecolor=darkblue,
  urlcolor=mediumblue]{hyperref}
\usepackage[capitalise,nameinlink]{cleveref}
\graphicspath{{/Users/takis/Downloads/}}

\definecolor{darkred}{rgb}{0.55,0.05,0.05}
\definecolor{darkblue}{rgb}{0.05,0.10,0.50}
\definecolor{mediumblue}{rgb}{0.05,0.25,0.65}
\definecolor{lightgray}{rgb}{0.94,0.94,0.94}
\definecolor{softblue}{rgb}{0.90,0.94,0.98}

\usepackage{adjustbox}

\journal{Computers \& Fluids}

\newcolumntype{Y}{>{\raggedright\arraybackslash}X}
\newcolumntype{C}{>{\centering\arraybackslash}X}

\lstdefinestyle{fortranstyle}{
  language=Fortran,
  basicstyle=\ttfamily\footnotesize,
  keywordstyle=\color{darkblue}\bfseries,
  commentstyle=\color{darkred},
  numbers=left,
  numberstyle=\tiny,
  stepnumber=1,
  numbersep=5pt,
  frame=single,
  rulecolor=\color{lightgray},
  breaklines=true,
  columns=fullflexible,
  showstringspaces=false
}

\newcommand{\ucns}{UCNS3D}
\newcommand{\archer}{ARCHER2}
\newcommand{\lumi}{LUMI}

\begin{document}

\begin{frontmatter}

\title{Towards Heterogeneous Exascale CFD with a Single Fortran Code Base:
OpenMP Target Acceleration of the High-Order Unstructured Solver UCNS3D}

\author[CU]{Panagiotis Tsoutsanis\corref{cor1}}
\ead{panagiotis.tsoutsanis@cranfield.ac.uk}
\author[CU]{Micha{\l} Umi\'nski}
\ead{Michal.Uminski@cranfield.ac.uk}
\author[EPCC]{Christopher Day}
\ead{c.day@epcc.ed.ac.uk}
\author[EPCC]{David Henty}
\ead{d.henty@epcc.ed.ac.uk}

\cortext[cor1]{Corresponding author}

\address[CU]{Faculty of Engineering and Applied Sciences, Cranfield University, Cranfield MK43 0AL, United Kingdom}
\address[EPCC]{EPCC, The University of Edinburgh, King's Buildings, Edinburgh EH9 3FD, United Kingdom}

\begin{abstract}
Heterogeneous exascale systems are reshaping computational fluid dynamics, but many mature high-order solvers cannot be rewritten for accelerators without fragmenting their code base, weakening reproducibility, or losing decades of numerical development. This paper presents a single-source OpenMP target modernisation of \ucns, a production high-order unstructured finite-volume solver for compressible flows. The work demonstrates that a complex scientific CFD code can be transformed into a common CPU/GPU implementation while preserving its numerical formulation, Fortran code base, and established production workflows.

The contribution is an end-to-end solver acceleration rather than a kernel-level port. The complete explicit time-advancement path is made device capable, including high-order reconstruction, gradient evaluation, inviscid and viscous flux integration, boundary treatment, halo exchange, and solution update. The implementation combines persistent OpenMP target data regions, a flat-array run-time layout, compile-time sizing of order-dependent temporary storage, explicit Fortran local algebra, target-callable routine restructuring, and GPU-aware MPI communication from device-resident buffers. These design choices address the core difficulty of high-order unstructured CFD on accelerators: irregular indirect stencils, medium arithmetic intensity, complex Fortran data structures, and large temporary storage requirements rather than dense linear-algebra workloads.

The accelerated solver is verified using the compressible Taylor--Green vortex, where CPU and GPU dissipation histories agree to machine precision on \lumi\ and remain consistent with published reference data. A single-node $\mathcal{P}_2$ Taylor--Green vortex run on the \archer\ GPU development platform gives the same machine-precision agreement and provides an independent portability check. Because the dominant irregular kernels do not expose the regularity required to approach hardware peaks, production performance is assessed end-to-end on fully populated CPU and GPU nodes using the Taylor–Green vortex, LM1021 sonic-boom, and NASA high-lift CRM benchmarks. Relative to the previous production CPU implementation, the refactored single-source CPU path is $(1.27-1.67\times)$ faster, while OpenMP target offload delivers same-node speed-ups of $(2.71-4.05\times)$ and retains (84-101\%) strong-scaling efficiency across the benchmark set. These results provide rigorous evidence that standards-based OpenMP can deliver portable, production-scale acceleration for mature high-order CFD solvers without sacrificing numerical fidelity, CPU performance, or long-term software sustainability.

\end{abstract}


\end{frontmatter}

\section{Introduction}
\label{sec:introduction}

High-order computational fluid dynamics (CFD) methods are attractive for transitional, vortical, aeroacoustic, and shock-turbulence dominated flows because they can deliver low numerical dissipation and high spectral resolution on comparatively coarse meshes. These benefits are particularly important for unstructured-mesh solvers, where complex engineering geometries must be represented without sacrificing accuracy in the resolved flow field. At the same time, high-order discretisations place substantial pressure on memory bandwidth, cache locality, communication patterns, and floating-point throughput. The cost of reconstruction, quadrature, and nonlinear limiting can dominate the run time, especially in three-dimensional simulations with mixed-element unstructured meshes.

The hardware landscape on which such solvers must operate has changed rapidly. Large CPU-only machines remain essential for production CFD, as exemplified by ARCHER2~\cite{beckett2024archer2}, but leadership-class systems increasingly combine many-core CPUs with GPUs or other accelerators. This shift is visible in European systems such as \lumi~\cite{lumi_hardware_2026}, 
JUPITER~\cite{jsc_jupiter_technical_2026,jupiter_user_docs_2026}, 
and HUNTER~\cite{hlrs_hunter_2026}, and it is expected to intensify as future heterogeneous exascale platforms are deployed. For research CFD codes, the central challenge is therefore not only to accelerate individual kernels, but to do so while preserving the scientific integrity, readability, and maintainability of the original solver.

Several open-source high-order CFD projects already demonstrate the value of accelerator-oriented algorithms on unstructured meshes. PyFR uses flux reconstruction with runtime code generation for streaming architectures \citep{pyfr}; Nektar++ provides a broad spectral/$hp$ element framework and has an ongoing performance-portability redesign for heterogeneous platforms \citep{Moxey2020Nektar,Renner2026NektarPerfPortable}; HORSES3D provides a high-order DG spectral-element multiphysics solver with a public GPU branch \citep{Ferrer2023HORSES3D,horses3dGpu}; and GAL{\AE}XI extends the FLEXI DGSEM line to accelerator-based systems, including large-GPU-count compressible-flow demonstrations \citep{Kurz2025GALAEXI}. These projects show that compact high-order element methods can map effectively to GPUs. The present work occupies a different niche: it starts from a mature Fortran, cell-centred, mixed-element finite-volume WENO/CWENO production code and keeps a single OpenMP/MPI source path, rather than adopting a domain-specific language, a C++ performance-portability layer, a CUDA/HIP/SYCL backend, or a separate GPU code branch.

Several programming routes are available. Vendor-specific approaches can expose excellent performance on a particular architecture, while portable abstraction layers can offer a more uniform application-facing interface. For mature Fortran solvers, however, each route has a different cost. Rewriting major parts of the solver in another language or maintaining separate back ends for different vendors can consume development effort that would otherwise be spent on numerical modelling. Directive-based OpenMP offers a different compromise: the original Fortran code remains largely intact, host parallelism and accelerator offload can be expressed through one standardised model, and architecture-specific details can be kept outside the numerical formulation where possible \citep{openmp52}.

This paper describes the modernisation of \ucns\ under the principle of ``one language, one directive model, one code base''. \ucns\ is a high-order unstructured CFD solver built around finite-volume discretisations for compressible flows and has been used for a range of smooth, shocked, and turbulent flow problems \citep{ANTONIADIS2022108453,Tsoutsanis_weno,Tsoutsanis_weno_viscous,Tsoutsanis_cwenos}. The present work focuses on the transition from the OpenMP CPU implementation in \ucns\ v3 to the OpenMP target implementation in \ucns\ v4. The objective is not to replace the numerical method, but to reorganise the run time so that the full explicit time-advancement path can execute on GPUs while preserving the CPU path, the Fortran code base, and the production workflows of the original solver.

This transition is more demanding than adding directives to a few expensive loops. The high-order methods increase the number of floating-point operations per cell, but the unstructured stencil access pattern prevents the kernels from behaving like dense linear-algebra workloads. Reconstruction stencils, quadrature loops, halo values, and boundary extrapolated states are gathered through indirect connectivity. In addition, the flexible derived-type data model used in the CPU solver is poorly matched to current OpenMP target compiler support, particularly when allocatable components are nested inside geometry, reconstruction, and solution containers. A useful GPU implementation therefore has to address memory layout, data lifetime, device communication, kernel call structure, and temporary storage together.

During implementation, we audited OpenMP accelerated-region syntax, target data clauses, variable mappings, and data-sharing attributes. Where used, automated code-review assistants supported source navigation, compiler-error tracking, and checklist-based inspection of device regions. Particular attention was given to variables referenced within device kernels, so that they were correctly mapped and scoped in explicit target data environments and \texttt{target teams distribute parallel do} constructs. This reduced the risk of compilation failures, unintended host-device transfers, stale device data, data races, and compiler-dependent behaviour. All findings and resulting changes were verified against the source code, compiler diagnostics, and regression tests.

The main contributions of this paper are:
\begin{itemize}[leftmargin=*]
  \item a structured methodology for the transition from \ucns\ v3 OpenMP CPU execution to the \ucns\ v4 OpenMP target GPU run time;
  \item a single-source implementation covering MPI+OpenMP execution on CPUs and OpenMP target offload on GPUs, while retaining Fortran as the only implementation language;
  \item an analysis of the dominant kernel transformations required for reconstruction, flux evaluation, gradients, time advancement, boundary treatments, and halo exchange;
  \item a description of the flat-array run-time layout, compile-time temporary sizing, GPU-aware MPI communication, and target-callable routine rewrites used to make the full explicit solver path device resident;
  \item a systematic audit of OpenMP accelerated-region syntax and target data clauses, supported by automated code-review assistants where useful;
  \item a performance and portability study on \archer\ and \lumi, including strong-scaling, parallel-efficiency, operation-profile, and GPU-acceleration results;
  \item a discussion of the practical limitations of directive-based GPU offload for complex Fortran CFD codes and the remaining path towards exascale readiness.
\end{itemize}

The remainder of the paper is organised as follows. \Cref{sec:solver} summarises the numerical and software structure of \ucns. \Cref{sec:strategy} presents the v3-to-v4 OpenMP target methodology and the code transformations required to support heterogeneous execution. \Cref{sec:systems} describes the target systems, build configurations, benchmark cases, and execution methodology. \Cref{sec:results} presents the verification and performance results. \Cref{sec:discussion} discusses portability, maintainability, and remaining bottlenecks, and \Cref{sec:conclusions} summarises the main findings.

\section{UCNS3D solver overview}
\label{sec:solver}

\subsection{Solver description}
\label{subsec:ucns_provenance}

\ucns\ is an open-source high-order finite-volume solver for compressible flows on arbitrary unstructured meshes \citep{ANTONIADIS2022108453}. Its numerical development includes mixed-element reconstruction, viscous high-order finite-volume schemes, explicit stencil-selection algorithms, central WENO/CWENOZ reconstruction, extended bounds limiting, and relaxed a posteriori MOOD limiting for shocked and multicomponent flows \citep{Tsoutsanis_weno,Tsoutsanis_weno_viscous,Tsoutsanis_stencil,Tsoutsanis_cwenos,TSOUTSANIS201869,TSOUTSANIS20218964,TSOUTSANIS2023127544,fluids9020033}. The solver lineage has also been used to study smooth and turbulent benchmark flows, atmospheric-flow formulations, low-Mach-number modifications, hypersonic shock-wave/turbulent-boundary-layer interaction, high-lift and RANS aerodynamics, rotorcraft and wind-turbine applications, and intake-flow distortion \citep{TSOUTSANIS2015207,Tsoutsanis_climate,Tsoutsanis_eccomas2016_lmc,Tsoutsanis_hyper,Tsoutsanis_hl,Tsoutsanis_rans,RICCI2020105648,SILVA2021106518,Silva2022,Silva2024,Tommaso2024dynamic}.

The code has therefore accumulated both numerical-method and application-specific infrastructure over many years. Previous optimisation studies on PRACE, ARCHER, and Knights Landing systems showed that \ucns\ performance is strongly influenced by memory layout, stencil traversal, and memory bandwidth \citep{Tsoutsanis_prace,Tsoutsanis_HOVE1,Tsoutsanis_HOVE2,Tsoutsanis_knl}. This history is important for the present work: the GPU port is not a new solver written around an accelerator-friendly discretisation, but a controlled modernisation of an established Fortran finite-volume code while retaining the solver capabilities and 1-to-1 exact numerical framework and validation record of the production branch.

\subsection{Governing equations and discretisation}
\label{subsec:governing}

The solver advances the compressible Navier-Stokes equations in conservative form,
\begin{equation}
  \frac{\partial \mathbf{U}}{\partial t}
  + \nabla \cdot \left(\vec{\mathbf{F}}_{c}(\mathbf{U})
  - \vec{\mathbf{F}}_{v}(\mathbf{U},\nabla\mathbf{U})\right)=0,
  \label{eq:ns}
\end{equation}
where $\mathbf{U}$ denotes the vector of conservative variables, and $\vec{\mathbf{F}}_{c}$ and $\vec{\mathbf{F}}_{v}$ denote the inviscid and viscous fluxes. The governing equations are integrated over an unstructured control volume $V_i$ to obtain
\begin{equation}
  \frac{d\mathbf{U}_i}{dt}
  =
  -\frac{1}{|V_i|}
  \sum_{f=1}^{N_f}
  \sum_{\alpha=1}^{N_q}
  \vec{\mathbf{F}}_{f,\alpha}
  \omega_{\alpha}|S_f|,
  \label{eq:fv_update}
\end{equation}
where $N_f$ is the number of faces of the element, $N_q$ is the number of quadrature points per face, $\omega_{\alpha}$ is the quadrature weight, and $|S_f|$ is the face measure. The numerical flux $\vec{\mathbf{F}}_{f,\alpha}$ is evaluated from left and right reconstructed states and, for viscous terms, reconstructed gradients.

\ucns\ supports arbitrary-order finite-volume reconstructions on mixed unstructured meshes consisting of any combination of conforming triangles, quadrilaterals, hexahedrals, pyramidals, tetrahedrals and prisms. The \ucns\ simulations reported in this work use the same numerical choices in all benchmark cases: least-squares polynomial reconstruction, CWENOZ reconstruction for non-oscillatory high-order state extrapolation, least-squares gradients for the viscous terms, the HLLC approximate Riemann solver for the inviscid fluxes, and an explicit fourth-order strong-stability-preserving Runge--Kutta time-integration scheme. Keeping this numerical configuration fixed is useful for the present OpenMP study because the CPU and GPU comparisons then isolate implementation effects rather than changes in the reconstruction method type.

The reconstruction machinery follows the unstructured high-order finite-volume formulation used in \ucns\ \citep{Tsoutsanis_weno,Tsoutsanis_weno_viscous,Titarev_cicp}. For a cell $i$, a polynomial of degree $r$ is expressed in a local reference coordinate system as
\begin{equation}
  \mathbf{p}_i(\xi,\eta,\zeta)
  =
  \mathbf{U}_i
  + \sum_{k=1}^{K} \mathbf{a}_{i,k}\phi_k(\xi,\eta,\zeta),
  \qquad
  K(r,d)=\frac{1}{d!}\prod_{\ell=1}^{d}(r+\ell),
  \label{eq:lsq_poly}
\end{equation}
where $d$ is the number of space dimensions, $\phi_k$ are zero-mean basis functions over the target cell, and $a_{i,k}$ are the polynomial degrees of freedom. The coefficients are obtained by requiring the reconstructed polynomial to preserve the cell averages over a compact stencil $\mathcal{S}_i$,
\begin{equation}
  \sum_{k=1}^{K} A_{mk}\mathbf{a}_{i,k}
  =
  |V_m|\left(\mathbf{U}_m-\mathbf{U}_i\right),
  \qquad
  A_{mk}=\int_{V_m}\phi_k\,dV,
  \quad V_m\in\mathcal{S}_i.
  \label{eq:lsq_system}
\end{equation}
This over-determined system is solved in a least-squares sense, with the required pseudo-inverse QR-based operators precomputed. Compact stencil-selection algorithms are used to keep the reconstruction robust and local on mixed-element unstructured meshes \citep{Dumbser2007204,Tsoutsanis_stencil}.

The high-order extrapolated states are obtained with the CWENOZ variant of the central weighted essentially non-oscillatory reconstruction implemented in UCNS3D \citep{ANTONIADIS2022108453,Tsoutsanis_cwenos}. This reconstruction follows the high-order unstructured finite-volume WENO framework developed for arbitrary mixed-element meshes \citep{Tsoutsanis_weno,Tsoutsanis_weno_viscous}, together with stencil-selection and robustness improvements for compact unstructured reconstructions \citep{Tsoutsanis_stencil,TSOUTSANIS201869}. The CWENO approach is particularly attractive on mixed-element unstructured grids because it preserves a compact central stencil while blending the optimal high-order polynomial with lower-order directional polynomials that provide non-oscillatory behaviour near discontinuities \citep{LevyPuppoRusso1999,Tsoutsanis_cwenos}.

For each control volume \(i\), an optimal polynomial \(\mathbf{p}^{\mathrm{opt}}_i\) is first constructed from the central stencil. This polynomial is then combined with lower-order directional polynomials \(\mathbf{p}_s\), built on biased sectorial stencils, through nonlinear CWENOZ weights. The reconstructed polynomial is written as
\begin{equation}
  \mathbf{p}_i^{\mathrm{CWENOZ}}(\xi,\eta,\zeta)
  =
  \sum_{s=1}^{s_t}\omega_s \mathbf{p}_s(\xi,\eta,\zeta),
  \qquad
  \omega_s=
  \frac{\tilde{\omega}_s}{\sum_{q=1}^{s_t}\tilde{\omega}_q},
  \label{eq:cwenoz_poly}
\end{equation}
where \(\lambda_s\) are the prescribed optimal linear weights and \(\omega_s\) are the nonlinear weights. The unnormalised weights are computed from smoothness indicators according to
\begin{equation}
  \tilde{\omega}_s
  =
  \lambda_s
  \left(1+\frac{\tau}{\epsilon+\mathcal{SI}_s}\right),
  \label{eq:cwenoz_weights}
\end{equation}
where \(\mathcal{SI}_s\) measures the oscillatory content of each candidate polynomial, \(\epsilon\) prevents division by zero, and \(\tau\) is a global smoothness measure following the WENO-Z philosophy \citep{BORGES20083191}.

\begin{figure*}[t]\begin{centering}

{\includegraphics[angle=0,width=0.32\textwidth]{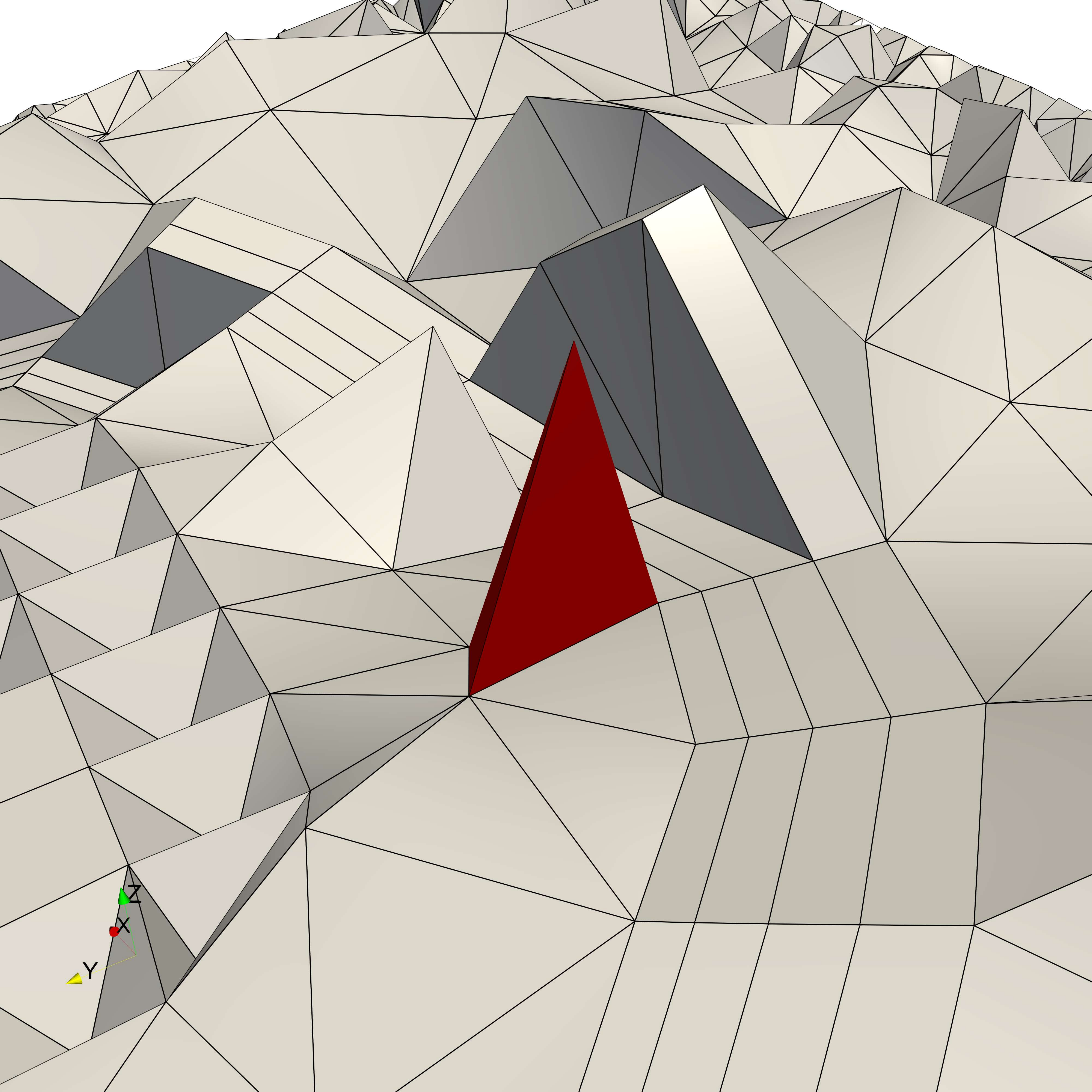}}
{\includegraphics[angle=0,width=0.32\textwidth]{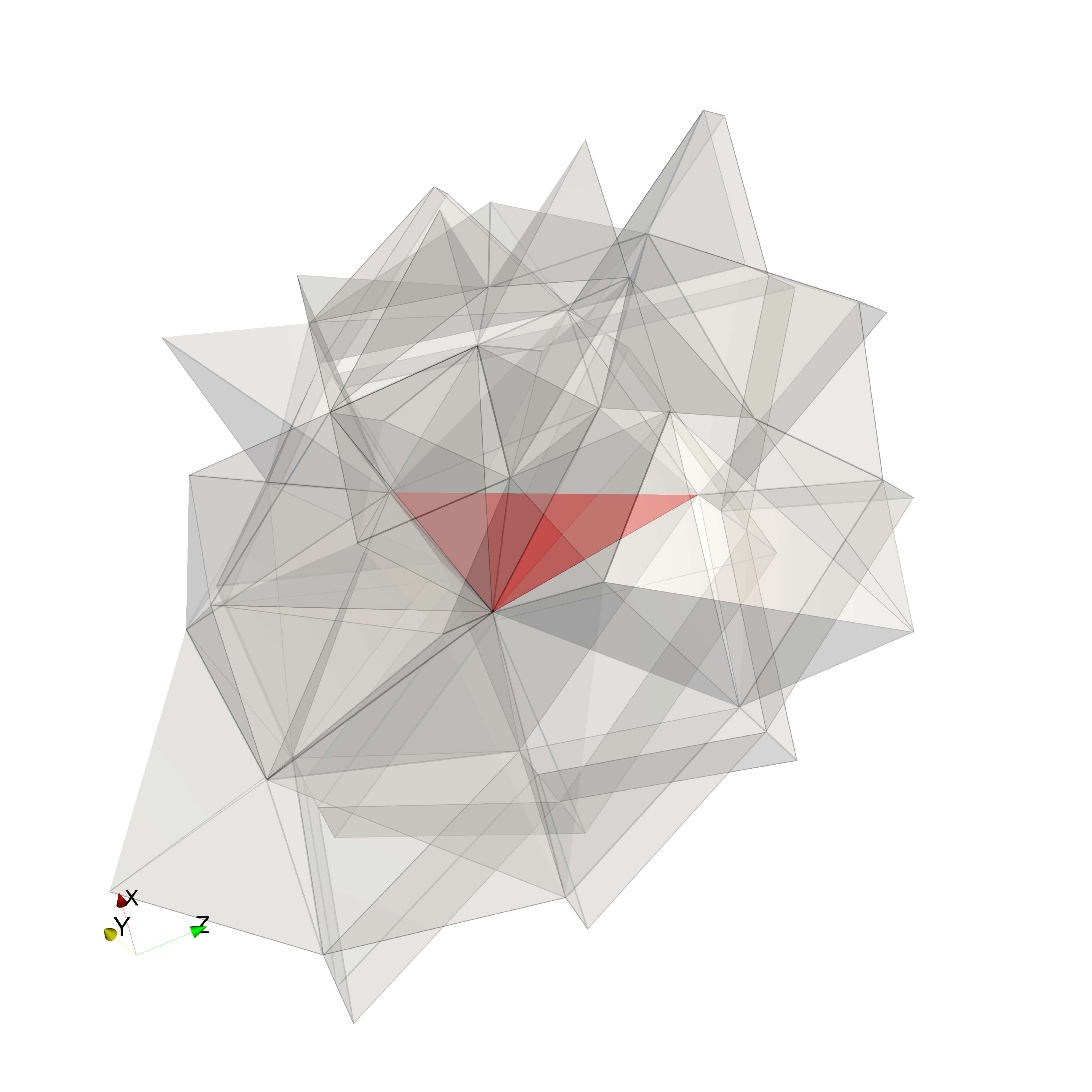}}
{\includegraphics[angle=0,width=0.32\textwidth]{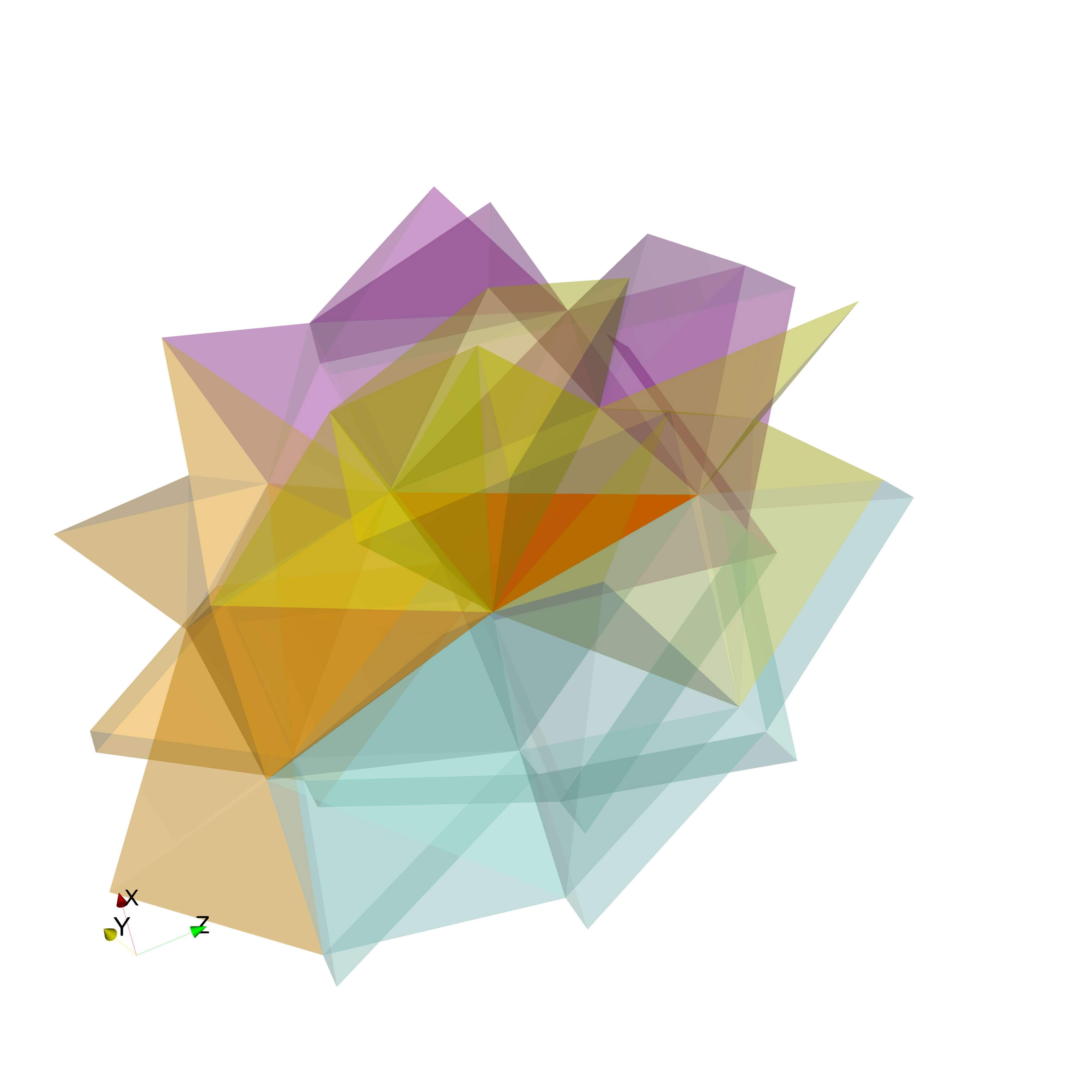}}
\par\end{centering}\caption{Considered cell with red colour (left), its central stencil elements  for the ($\mathcal{P}_{4}$) polynomial (middle), and the directional stencil elements for the lower-order ($\mathcal{P}_{2}$) polynomials (right), each one of them illustrated with different colour for a CWENOZ5 ($\mathcal{P}_{4}$) scheme.}\label{fig:stencils}\end{figure*}

In the present UCNS3D implementation, the degree of the lower-order directional polynomials is selected according to the target order of the CWENOZ reconstruction. For the third- and fourth-order CWENOZ schemes, the biased directional stencils use linear polynomials, which provide a robust second-order accurate lower-order reconstruction. For fifth-order CWENOZ schemes and higher, the directional stencil polynomials are quadratic, providing a third-order accurate lower-order reconstruction as shown in \Cref{fig:stencils}. This choice keeps the sectorial substencils compact and robust near discontinuities, while allowing the central optimal polynomial to recover the prescribed high-order accuracy in smooth regions.

In smooth regions, the smoothness indicators of the candidate polynomials are comparable and the nonlinear weights tend to the optimal linear weights. The reconstruction therefore recovers the designed high-order central polynomial. Near shocks, contact discontinuities, material interfaces, or under-resolved turbulent gradients, the smoothness indicators penalise oscillatory stencil contributions and shift the reconstruction toward smoother directional polynomials. This reduces spurious oscillations while retaining the compact stencil footprint required for efficient high-order finite-volume computations on arbitrary hybrid unstructured meshes.

The resulting CWENOZ strategy preserves the accuracy and compactness required by UCNS3D for compressible simulations involving shocks, turbulence, and complex geometries \citep{Tsoutsanis_prace,Tsoutsanis_rans}. The same reconstruction philosophy has also been extended in previous work to multicomponent and interface-capturing applications, where compact CWENO reconstructions are useful for maintaining sharp but stable material interfaces on unstructured meshes \citep{TSOUTSANIS20218964,TSOUTSANIS2023127544,DG_FV_MULTI}.

The viscous fluxes use gradients reconstructed by the least-squares procedure, with boundary-condition constraints applied where required \citep{Tsoutsanis_weno_viscous}. At an interface, the left and right gradients used in the viscous stress tensor and heat flux are averaged with a jump penalty,
\begin{equation}
  \nabla\mathbf{U}_f
  =
  \frac{1}{2}
  \left(\nabla\mathbf{U}_L+\nabla\mathbf{U}_R\right)
  + \frac{\alpha}{L_{\mathrm{int}}}
  \left(\mathbf{U}_R-\mathbf{U}_L\right)\otimes\vec{n}_f,
  \label{eq:viscous_gradient_penalty}
\end{equation}
where $L_{\mathrm{int}}$ is the distance between neighbouring cell centres, $\vec{n}_f$ is the interface normal, and $\alpha=4/3$ follows the penalty treatment used in the baseline solver.

The inviscid numerical flux at each face quadrature point is evaluated with the HLLC approximate Riemann solver of Toro et al. \citep{HLLC}. The resulting semi-discrete system,
\begin{equation}
  \frac{d\mathbf{U}_i}{dt}=\mathbf{R}_i(\mathbf{U}),
  \label{eq:semi_discrete}
\end{equation}
is advanced with the explicit fourth-order SSP Runge--Kutta method available in \ucns\ \citep{Gottlieb199873,SSP4}. The explicit Runge--Kutta stages repeatedly execute the same reconstruction, gradient, flux-integration, residual-accumulation, and update kernels, which is why these methods define the main computational workload targeted by the OpenMP CPU and GPU implementations.

\subsection{Baseline software structure}
\label{subsec:software}

The original solver is a Fortran MPI code in which the mesh is partitioned across ranks. Each rank owns a set of local cells and stores halo data for neighbouring partitions. The major modules include mesh and topology handling, reconstruction, gradient evaluation, Riemann solvers, viscous fluxes, boundary conditions, time integration, and input/output. A simplified view of the time-step loop is shown in \Cref{alg:timestep}.

\begin{algorithm}[t]
\caption{Representative \ucns\ time-step structure}
\label{alg:timestep}
\begin{algorithmic}[1]
\For{Runge-Kutta stage $s=1,\ldots,N_s$}
  \State Exchange halo values through MPI
  \State Reconstruct conservative or primitive variables
  \State Evaluate gradients and boundary states
  \State Exchange inter-processor gradients and boundary states
  \State Integrate inviscid and viscous fluxes at faces
  \State Accumulate residuals in each control volume
  \State Update the conservative solution
\EndFor
\end{algorithmic}
\end{algorithm}

Profiling of representative high-order cases shows that the expensive regions are dominated by local numerical kernels rather than by a single monolithic operation, as detailed in \citep{Tsoutsanis_prace}. In particular, reconstruction and flux integration are both arithmetic intensive and repeatedly traverse indirect mesh connectivity. In the existing hybrid MPI+OpenMP implementation, these kernels are parallelised on multicore CPUs; the present work extends their execution to GPUs using OpenMP target offloading. Achieving efficient GPU performance requires careful treatment of data residency and movement, memory layout, loop organisation, temporary storage, and indirect-access patterns, together with sufficient parallelism for the compiler to generate efficient device code.


\section{Methodology for the v3-to-v4 OpenMP target transition}
\label{sec:strategy}

\subsection{Design principles}
\label{subsec:principles}

The v3-to-v4 transition was organised as a run-time modernisation rather than as a numerical-method rewrite. The numerical fluxes, reconstruction choices, time integration, and verification cases were preserved, while the storage model and execution model were changed so that the same source could support CPU execution and GPU offload. This distinction is important: for a high-order unstructured solver, the main difficulty is not expressing a single loop with \texttt{target teams distribute parallel do}, but making sure that all data and all routines reached by the time-step loop are valid, efficient, and resident on the target device.

The porting strategy is built around four pillars. First, Fortran remains the implementation language. This retains compatibility with the existing solver, avoids a wholesale rewrite, and allows the developers to continue working in the language in which the code base is already expressed. Second, OpenMP is used as the only node-level programming model. CPU threading and accelerator offload are therefore represented by the same directive family. Third, the same source files are used for CPU-only and GPU-enabled builds, with conditional compilation limited to build-system choices, compiler workarounds, and small target-specific tuning parameters. Fourth, the modernisation is treated as a controlled engineering process. Each target region is checked for performance, correct data use, device compatibility, and numerical agreement with the CPU baseline.

The solver already uses a hybrid MPI+OpenMP programming model on CPU systems, as demonstrated in earlier UCNS3D performance work \citep{Tsoutsanis_prace}. The present work therefore preserves the existing MPI-based domain decomposition and extends the node-level OpenMP parallelism to accelerators through OpenMP target offload. On CPUs, each MPI rank can use multiple OpenMP threads, while on GPU nodes MPI ranks are associated with accelerators and the most expensive loops are executed on the device. Previous Knights Landing results, where UCNS3D benefited strongly from the higher-bandwidth MCDRAM memory, also indicate that the solver is sensitive to memory bandwidth and memory latency, largely because high-order unstructured stencils involve indirect and non-coalesced memory access patterns.

\begin{table*}[t]
\centering
\caption{Principal challenges in the transition from the \ucns\ v3 OpenMP CPU implementation to the \ucns\ v4 OpenMP target GPU implementation, and the corresponding methodology adopted in the port.}
\label{tab:transition_challenges}
\small
\begin{tabularx}{\textwidth}{@{}p{0.29\textwidth}p{0.32\textwidth}p{0.32\textwidth}@{}}
\toprule
Challenge & Effect on the GPU port & Selected approach in \ucns\ v4 \\
\midrule
Medium arithmetic intensity with irregular unstructured stencils & High-order reconstruction increases work per cell, but indirect gathers from stencil elements limit locality and memory coalescing. The kernels do not map naturally to dense matrix operations. & Preserve cell- and face-parallel work sharing, flatten stencil connectivity and reconstruction data, keep inner loops explicit, and minimise repeated indirect loads inside quadrature and polynomial loops. \\
Full run time must execute on the device & Porting only reconstruction or flux kernels would leave expensive host-device transfers around every Runge-Kutta stage and halo exchange. & Use persistent OpenMP target data regions around the explicit time loop. Update host data only for output, restart, and time step synchronisation. \\
Derived types with allocatable components & The v3 data model is flexible for mesh, geometry, reconstruction, and solution storage, but nested allocatable components have inconsistent and fragile OpenMP target support across compilers. & Use flat run-time arrays with explicit offsets for frequently accessed data. This reduced memory footprint by approximately two to six times in representative paths and gave compilers simpler device mappings. \\
Small matrix operations in reconstruction & Calling vendor BLAS libraries would introduce extra dependencies and architecture-specific tuning, while the matrices are often small, case dependent, and embedded inside irregular loops. & Retain Fortran-only explicit loops for matrix-matrix and matrix-vector operations. Micro-kernel tests showed these loops to be competitive on the relevant compiler stacks while remaining portable. \\
Large temporary arrays and deep call chains & Target kernels that call several layers of subroutines can create high register pressure, spilling, and reduced occupancy, especially when each level introduces private work arrays. & Determine maximum temporary sizes at build time from a generic or case-specialised \texttt{UCNS3D\allowbreak.DAT} profile, make work-array dimensions compile-time constants, and split or simplify device-call paths where necessary. \\
Functions returning arrays from device kernels & Several OpenMP target compiler stacks do not reliably support array-valued function returns from device code, particularly inside nested Fortran call chains. & Replace array-valued functions by scalar functions or by subroutines that write into caller-owned arrays. This also makes data scoping and temporary ownership explicit. \\
MPI communication of halo and boundary data & Host staging of halo values would reintroduce the data-transfer overhead that persistent device residency is designed to avoid. & Use GPU-aware MPI point-to-point communication with device addresses for stencil halo buffers and interprocessor boundary extrapolated values at quadrature points, with host staging retained only as a fallback path. \\
\bottomrule
\end{tabularx}
\end{table*}

\subsection{Device-resident data lifetime and flat run-time layout}
\label{subsec:data}

The first requirement for OpenMP target offload is to make data lifetime explicit. In the baseline CPU code, allocatable arrays and derived types are created, filled, and accessed through normal host memory semantics. On a device, repeated implicit transfers can dominate run time and obscure the computational benefit of offload. The v4 implementation therefore moves from a host-centred execution model to a device-resident execution model: mesh geometry, connectivity, quadrature data, reconstruction coefficients, solution arrays, residual arrays, boundary buffers, and MPI halo buffers are created or transferred to the device before the explicit time loop and remain there across Runge-Kutta stages.

This required a change in the run-time representation of the solver. The v3 code uses derived types with allocatable components to express cell geometries, reconstruction stencils, solution states, and auxiliary data. This representation is convenient and flexible for CPU development, but it is poorly suited to current OpenMP target implementations because each allocatable component has its own descriptor, mapping state, and compiler-dependent device semantics. In v4, the data used by accelerated kernels are therefore stored in flat arrays with explicit offsets and bounds. The numerical meaning of the data is unchanged, but the memory model becomes simple enough for OpenMP target mapping, device pointer communication, and compiler analysis. This flattening also removes descriptor and indexing overhead associated with many small allocatable components; in representative run-time paths the memory footprint was reduced by approximately two to six times depending on the polynomial order.

The typical execution pattern is illustrated in \Cref{lst:targetdata}. The important feature is that the time integration loop does not repeatedly move the conservative solution, residual, geometry, reconstruction, or halo buffers between host and device. Host updates are limited to output, restart files, selected diagnostics, and fallback communication paths.

\begin{lstlisting}[style=fortranstyle,
caption={Schematic persistent OpenMP target data pattern used for the accelerator execution path.},
label={lst:targetdata}]
! Allocate persistent device storage and initialise it explicitly.
call initialise_device_data()
! Internally:
!$omp target enter data map(alloc: mesh_data, solution_data, rhs_data)
!$omp target enter data map(alloc: reconstruction_data, halo_buffers)
!$omp target update to(mesh_data, solution_data, reconstruction_data)

! Keep the main time-advancement data live on the device.
!$omp target data map(to: run_control, mesh_data, reconstruction_data, &
!$omp& solution_data, rhs_data, halo_buffers, boundary_buffers)

do while (.not. finished)
  call compute_time_step()
  !$omp target update to(dt)
  call exchange_halos_device_aware()
  call reconstruct_solution_on_device()
  call exchange_interprocessor_boundaries_device_aware()
  call compute_fluxes_on_device()
  !$omp target teams distribute parallel do &
  !$omp& map(alloc: solution_data, rhs_data, cell_volume)
  do i = 1, nlocal_cells
    call update_solution_stage(i, dt)
  end do
  !$omp end target teams distribute parallel do
  if (output_step) then
    !$omp target update from(solution_data)
    call write_solution_on_host()
  end if
end do
!$omp end target data
\end{lstlisting}

The target data environment also has to be complete. In a large Fortran CFD code, it is easy for a variable to appear inside an offloaded loop because it is available through a module, host association, or a deep call chain, while not being explicitly accounted for in the target data region. Such loose variables can lead to implicit transfers, host fallbacks, stale device data, or compiler-dependent behaviour. The port therefore treats target-clause completeness as a correctness requirement. In the accelerated kernels, scalar values are passed using \texttt{firstprivate}, while arrays that have already been allocated and initialised on the device are referenced with \texttt{map(alloc:\ldots)} rather than \texttt{present}, for better compiler portability. Explicit \texttt{target update} directives are used only where host-device synchronisation is required, for example for time-step control, diagnostics, output, or termination flags.

\subsection{Compile-time sizing and target-callable routines}
\label{subsec:compile_time}

A second methodology choice was to move as many temporary-array decisions as possible from run time to build time. High-order reconstruction, CWENO/WENO weighting, viscous flux evaluation, turbulence and passive-scalar extensions, and real-gas source terms all require local work arrays. If these arrays are allocated inside target kernels, or if their size is hidden behind a long call chain, the compiler has little opportunity to place them efficiently. Conversely, if every possible configuration is compiled into a single generic GPU binary, private temporary storage can become excessively large and register pressure can limit occupancy.

The v4 build process therefore follows a case-specialised compilation philosophy. A Python preprocessing step reads either a generic \texttt{UCNS3D\allowbreak.DAT} profile or the specialised \texttt{UCNS3D\allowbreak.DAT} file for the intended GPU simulation and extracts maximum values for dimensions such as polynomial order, stencil type, number of variables, number of species, turbulence variables, passive scalars, degrees of freedom, neighbours, and face or volume quadrature points. These limits are compiled as parameters for the GPU build. The approach is similar in philosophy to just-in-time specialisation at the case-preparation stage: the source remains general, but the compiler sees the upper bounds for the temporary arrays used by the selected simulation class.

This choice serves two purposes. First, it makes device-private storage explicit and auditable, which is essential for OpenMP target correctness. Second, it exposes the pressure points to the compiler and to the developer: if a kernel requires too many registers or too much private memory for a particular GPU, the problematic temporary arrays and call paths can be identified directly. In practice, this led to routine-level changes such as reducing the scope of temporary variables, splitting overly large kernels, and replacing hidden array temporaries with caller-owned work arrays.

The call structure also had to be made target compatible. Several Fortran routines in the CPU implementation returned small arrays from functions, a style that is expressive on the host but not reliably supported inside OpenMP target kernels by all relevant compiler stacks. These routines were rewritten either as scalar functions, when only one value is required, or as subroutines that write into an explicit output array supplied by the caller. This change also reduces hidden allocations and clarifies which temporary data are private to a cell, face, quadrature point, or Runge-Kutta stage.

The same reasoning was applied to local matrix operations. The reconstruction machinery contains many small matrix-vector and matrix-matrix products, but they are embedded in irregular unstructured loops and their dimensions depend on the selected order and stencil. Vendor BLAS libraries can be effective for large, dense operations, but relying on them here would introduce additional dependencies and reduce portability. Simple Fortran loops were therefore retained and specialised by the compiler. Micro-kernel tests with recent Cray, Nvidia, and AMD compiler stacks showed that these explicit loops were competitive for the small operations of interest while keeping the code Fortran-only and single-source.

\subsection{Loop restructuring and kernel granularity}
\label{subsec:loops}

High-order unstructured CFD kernels contain nested loops over elements, faces, quadrature points, variables, stencil entries, and polynomial modes. Although high-order methods increase the number of operations per cell, the arithmetic intensity of the production kernels remains only moderate because much of the work is driven by irregular stencil gathers and indirect connectivity. The GPU implementation therefore cannot rely on the memory-access regularity or data reuse typical of dense tensor or BLAS-like workloads. Its performance depends instead on exposing enough cell- or face-level parallelism, keeping the inner loops simple, and limiting the number of temporary values live at the same time.

On CPUs, the outer cell or face loop provides a natural OpenMP parallel loop. On GPUs, the same loop must expose enough parallelism to occupy the device while avoiding excessive register pressure and redundant memory traffic. The port therefore separates large independent loops from small serial reductions, keeps gather patterns explicit, and limits fusion when a larger fused kernel would carry too many local arrays through a deep call chain. This was an important practical compromise: fewer, larger kernels can reduce launch overhead, but in reconstruction and flux integration they can also increase register use, trigger spilling, and reduce occupancy.

For CPU builds, the dominant pattern is a cell- or face-parallel
\texttt{parallel do}. For GPU builds, the corresponding pattern is a
cell- or face-parallel \texttt{target teams distribute parallel do}.
Temporary arrays are made private to each iteration where possible. Race conditions in residual accumulation are avoided by selecting ownership-based loop structures or by separating face-based flux evaluation from cell-based residual update. 

\begin{lstlisting}[style=fortranstyle,caption={Representative single-source loop form for CPU threading and GPU offload.},label={lst:loop}]
!$omp target teams distribute parallel do collapse(1) &
!$omp& firstprivate(ncells) map(alloc: q, rhs, geometry, connectivity)
do i = 1, ncells
  call reconstruct_cell(i, q, geometry, connectivity, scratch)
  call integrate_cell_flux(i, q, rhs, geometry, scratch)
end do
\end{lstlisting}

\subsection{Accelerated-region syntax and target-clause audits}
\label{subsec:llm_debugging}

Because accelerator-enabled builds of the full Fortran code base were substantially more time consuming than CPU-only builds, typically by a factor of 4-10, candidate OpenMP target regions were audited before full compilation. The checks focused on directive syntax, variable scoping, data-mapping clauses, reductions, and the device availability of routines reached from offloaded kernels. This was particularly important in a large Fortran code base, where variables and callees may be introduced through modules, host association, contained procedures, or deeper call chains that are not apparent from an isolated loop nest.
The audit combined compiler diagnostics, manual source inspection, and targeted runtime tests. Automated code-review assistants were used only as an additional screening aid to produce checklist-style reviews of long OpenMP target constructs and to flag possible scoping or data-mapping omissions. All candidate issues were verified against the source code, compiler output, and runtime behaviour before any change was accepted.

\subsection{Ported kernels}
\label{subsec:kernels}

In version 4, the GPU port targets the complete explicit execution path used for production calculations, rather than isolated kernels only. The offloaded workflow includes the spatial discretisation machinery from first to seventh order, covering the linear, MUSCL, (C)WENO, and MOOD reconstruction options available in the production solver. It also includes the inviscid and viscous flux evaluation paths required for Euler and Navier-Stokes calculations under explicit time stepping, including iLES-type simulations.
Implicit solvers and algorithms dominated by strong global coupling were not included in this first GPU port. The explicit time-marching loop was selected as the initial target because it provides a more direct route to sustained device residency, exposes substantial element- and face-level concurrency, and requires comparatively limited synchronisation between kernels and MPI exchanges. Consequently, the reported timings correspond to complete explicit solver iterations for the selected production paths, not only to standalone reconstruction or flux kernels. \Cref{tab:kernels} summarises the principal kernels and the changes made during modernisation.

\begin{table*}[t]
\centering
\caption{Principal \ucns\ kernels ported or restructured during the v3-to-v4 OpenMP target modernisation.}
\label{tab:kernels}
\begin{tabularx}{\textwidth}{@{}p{0.20\textwidth}p{0.28\textwidth}p{0.31\textwidth}p{0.13\textwidth}@{}}
\toprule
Kernel & Dominant operation & OpenMP modernisation & Status \\
\midrule
Polynomial reconstruction & Stencil gather, least-squares coefficients, nonlinear weights, state evaluation & Flattened stencil access, compile-time work-array bounds, private temporaries, explicit local algebra & Ported in v4 \\
Gradient evaluation & Cell and face gradients for viscous fluxes & Cell-parallel loops, explicit mapping of geometry and state arrays, target-callable helpers & Ported in v4 \\
Inviscid flux integration & Riemann solves at face quadrature points & Face/cell parallel kernels, reduced temporary allocation, scalar-return helper functions & Ported in v4 \\
Viscous flux integration & Gradient-dependent fluxes and transport properties & Device-capable routines, data reuse within quadrature loops, flattened geometry access & Ported in v4 \\
Boundary conditions & Ghost state construction and wall/inflow/outflow treatments & Separated boundary sets, device-resident boundary buffers, target-capable helper routines & Ported in v4 \\
Time update & Runge-Kutta state updates and residual scaling & Bandwidth-bound target kernels with persistent solution and residual arrays & Ported in v4 \\
MPI halo and boundary exchange & Stencil values and interprocessor extrapolated states & Device pack/unpack kernels and GPU-aware MPI point-to-point communication & Ported in v4 \\
Target-clause audits & Data scoping and device availability across offloaded call chains & Manual and automated checks for missing or wrongly scoped target declarations & Continuous \\
\bottomrule
\end{tabularx}
\end{table*}

\subsection{GPU-aware MPI communication and accelerator placement}
\label{subsec:placement}

The modernised code retains MPI for distributed-memory parallelism. This was a deliberate choice: the mesh partitioning, halo ownership, and interprocessor boundary logic are already mature in the v3 solver, and replacing MPI would create unnecessary risk. The v4 methodology instead keeps the distributed-memory structure and changes the location of the exchanged data. Halo values for reconstruction stencils and interprocessor boundary extrapolated states at face quadrature points are packed on the device, communicated using GPU-aware MPI where available, and unpacked on the device. The MPI calls are given the device addresses of the send and receive buffers, avoiding the host staging that would otherwise occur at every Runge-Kutta stage.

This communication path is essential for an end-to-end GPU run. If only the numerical kernels are offloaded while halo exchange is staged through the host, the solver pays repeated device-host and host-device transfers around exactly the parts of the algorithm that are executed most frequently. By contrast, direct device-buffer communication keeps the principal solution, residual, geometry, reconstruction, halo, and boundary data resident on the GPU\@. Host copies are reserved for output, restart, diagnostics, and fallback paths on systems where GPU-aware MPI is not reliable for the selected compiler and MPI stack.

\section{Systems and methodology}
\label{sec:systems}

\subsection{Target platforms}
\label{subsec:platforms}

The experimental programme uses \lumi\ as the primary performance-measurement platform. The \archer\ GPU development platform is used for code development and compiler/runtime testing, and it provides an independent single-node $\mathcal{P}_2$ Taylor--Green vortex reproducibility check. The benchmark-wide production timings reported below are from \lumi. The \lumi\ CPU and GPU runs enable direct comparison between the CPU and GPU builds within the same system environment. The \lumi\ CPU runs provide a reference for the v4 CPU implementation, while the \lumi\ GPU runs evaluate the OpenMP target implementation, persistent device data strategy, and GPU-aware MPI communication. The hardware, compiler, MPI, and OpenMP runtime versions are reported explicitly, since these details can strongly affect both CPU threading and OpenMP target performance.

The performance methodology compares three solver configurations. The first is the \ucns\ v3 CPU implementation, which provides the mature MPI+OpenMP reference for time to solution and numerical behaviour. The second is the \ucns\ v4 CPU build, which isolates the effect of the flat run-time layout and OpenMP restructuring without GPU offload. The third is the \ucns\ v4 GPU build, which adds OpenMP target execution, persistent device data, and GPU-aware MPI communication. This separation is important because any speed-up of the v4 GPU build over a CPU baseline combines several effects: data-layout changes, CPU-side OpenMP restructuring, accelerator offload, and reduced memory footprint.

\begin{table}[t]
\centering
\scriptsize
\setlength{\tabcolsep}{2.2pt}
\renewcommand{\arraystretch}{1.12}
\caption{System configuration used for the \archer\ GPU development check and the \lumi\ CPU/GPU performance measurements.}
\label{tab:systems}
\begin{adjustbox}{max width=\columnwidth}
\begin{tabularx}{\columnwidth}{@{}p{0.22\columnwidth}YY@{}}
\toprule
Item & \archer & \lumi \\
\midrule
Node type
& GPU development node
& LUMI-G GPU node \\

Processor
& 1$\times$ AMD EPYC 7543P, 32 cores, 2.8 GHz
& 1$\times$ AMD EPYC 7A53 ``Trento'', 64 cores \\

Accelerator
& 4$\times$ AMD Instinct MI210
& 4$\times$ AMD Instinct MI250X;\newline 8 GCDs visible\\

Memory per node
& 512 GiB host; 64 GiB HBM2e per GPU
& 512 GiB host; 128 GiB HBM2e per MI250X \\

Interconnect
& HPE Cray Slingshot; 2$\times$100 Gb/s interfaces
& HPE Cray Slingshot-11; 4 endpoints per GPU node \\

Compiler
& Cray PE 25.09 with ROCm 6.3.4
& Cray PE 25.03 with ROCm 6.3.4 \\

MPI library
& \multicolumn{2}{c}{Cray MPICH, GPU-aware MPI enabled}\\

OpenMP support
& \multicolumn{2}{c}{OpenMP target offload through CCE/ROCm} \\
\bottomrule
\end{tabularx}
\end{adjustbox}
\end{table}

\subsection{Benchmark cases}
\label{subsec:benchmarks}

The benchmark suite was selected to assess both numerical reproducibility across execution modes and computational performance. The Taylor--Green vortex (TGV) cases provide a controlled test for comparing accelerated and non-accelerated compilations, using identical numerical settings for CPU-only and GPU-offloaded runs. These cases exercise the high-order reconstruction, face quadrature, and turbulence-resolution components of the solver on structured hexahedral meshes. The LM1021 and NASA high-lift CRM cases then extend the assessment to progressively more representative aerodynamic configurations, involving hybrid unstructured meshes, complex geometries, strong gradients, and production-scale memory footprints. \Cref{tab:benchmarks} reports the mesh size, the estimated number of unique mesh faces, the total reconstructed degrees of freedom per flow variable, and the total number of face quadrature points. The latter also gives the approximate number of Riemann problems solved during each residual evaluation; by comparison, a standard second-order finite-volume scheme solves approximately one Riemann problem per mesh face.

\begin{figure*}[t]\begin{centering}

{\includegraphics[angle=0,width=0.33\textwidth]{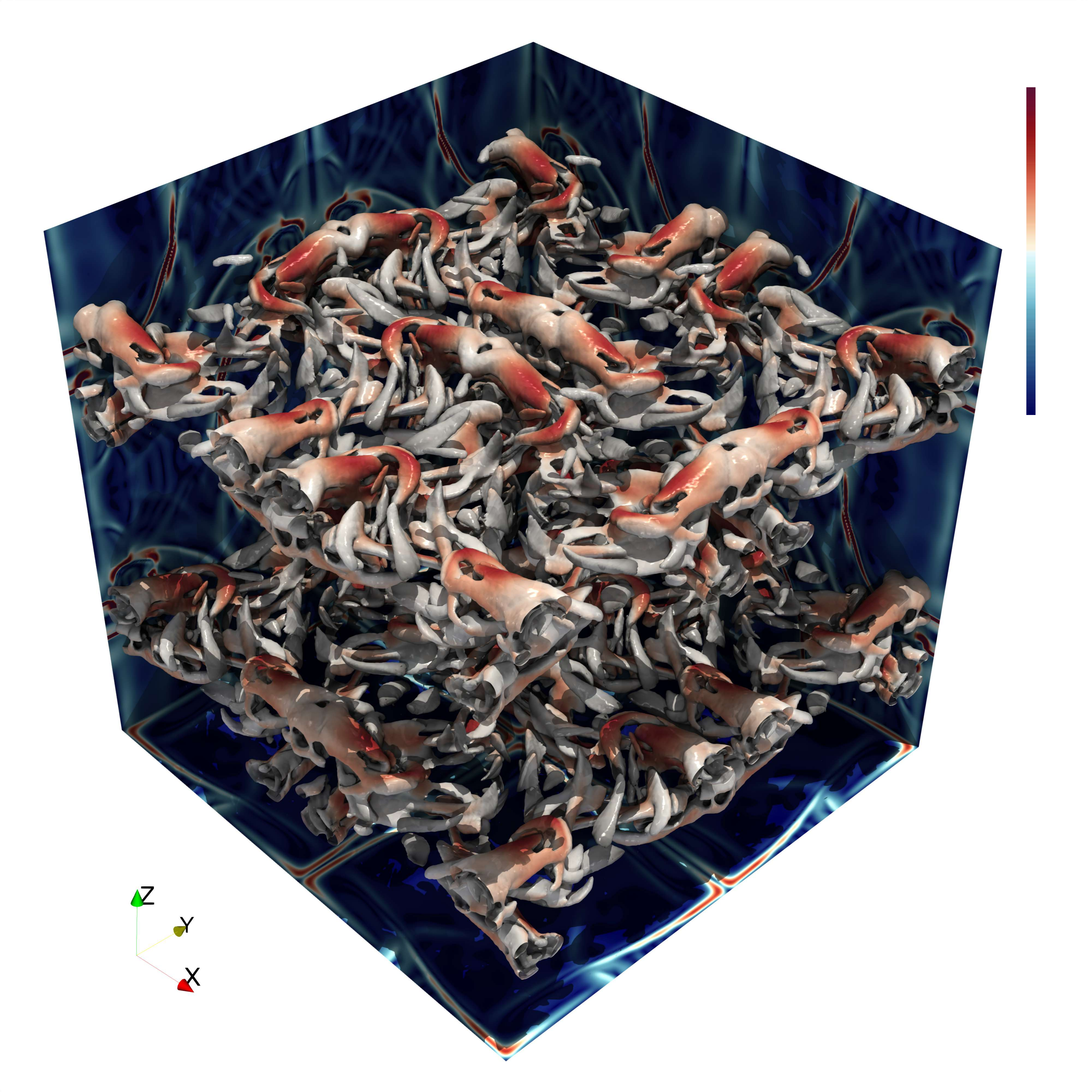}}
{\includegraphics[angle=0,width=0.33\textwidth]{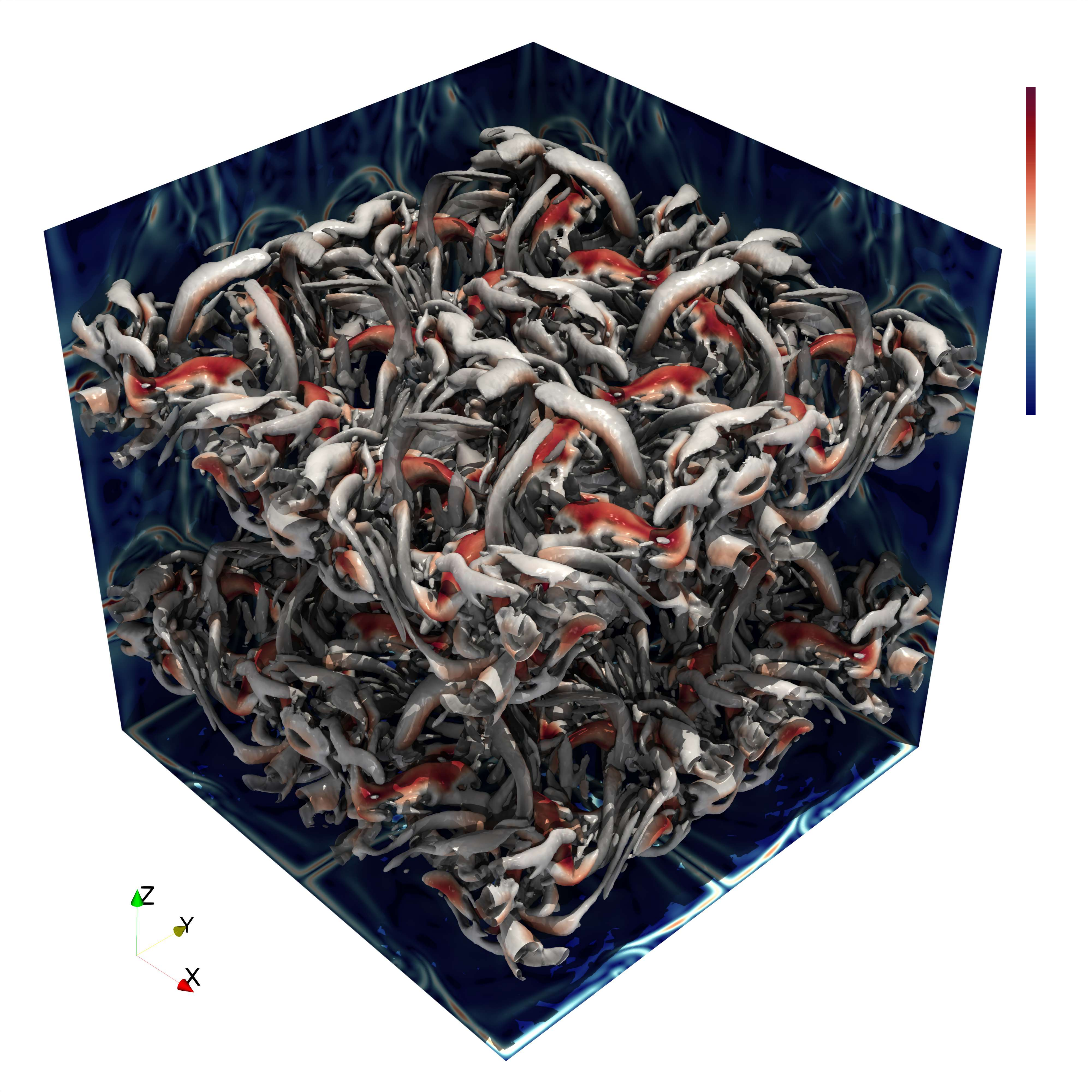}}
{\includegraphics[angle=0,width=0.33\textwidth]{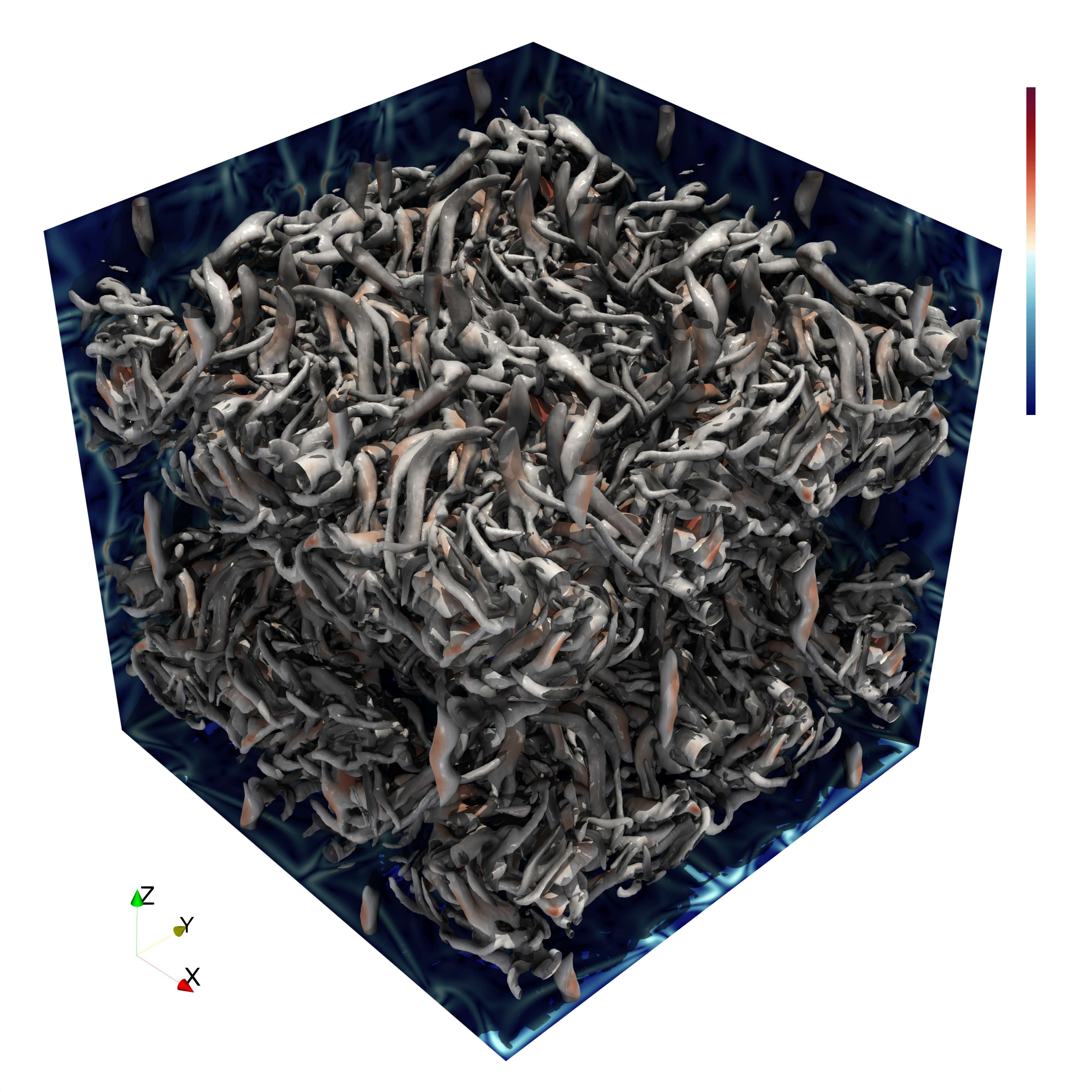}}
\par\end{centering}\caption{Taylor--Green vortex isosurfaces of Q-criterion coloured by Mach number and contour plots of density-gradient magnitude at the $x=\pi$, $y=\pi$, and $z=\pi$ planes obtained with the CWENOZ5 scheme at $t=8$ (left), $t=9$ (middle), and $t=11$ (right).}\label{fig:tgv_q_isosurfaces}\end{figure*}

\paragraph{\textbf{Supersonic viscous Taylor--Green vortex}}
The supersonic viscous TGV is used as a compact compressible-turbulence benchmark because it combines smooth vortical evolution, shocklet formation, dilatational dissipation, and sensitivity to numerical dissipation in a periodic box, as illustrated by the UCNS3D visualisation in \Cref{fig:tgv_q_isosurfaces}. The case follows the compressible TGV variant introduced by Lusher and Sandham \citep{tgv_super} and subsequently used for cross-code high-order comparisons by Chapelier et al. \citep{Chapelier2024}. The computational domain is a periodic box of side length $2\pi$ in each coordinate direction. The initial velocity field is
\begin{alignat}{3}\label{eq:TaylorGreen-velocity}
u(x,y,z,0) &=  &&\sin(x)\cos(y)\cos(z)\nonumber\\
v(x,y,z,0) &= -&&\cos(x)\sin(y)\cos(z)\\
w(x,y,z,0) &=  &&0\nonumber
\end{alignat}
with pressure
\begin{equation}\label{eq:TaylorGreen-pressure}
p(x,y,z,0)=\frac{1}{\gamma M_{\mathrm{ref}}^2}
  +\frac{1}{16}\left[\cos(2z)+2\right]\left[\cos(2x)+\cos(2y)\right]
\end{equation}
and the density obtained from the equation of state with initially uniform temperature. The performance runs use the $M_{\mathrm{ref}}=1.25$, $Re=1600$ viscous configuration on a $128^3$ hexahedral mesh, corresponding to approximately 2.0 million cells. The principal diagnostics are the evolution of normalised kinetic energy and its solenoidal and dilatational dissipation components; these quantities are compared with the reference high-order TENO, DG, and multi-code datasets reported in \citep{tgv_super,Chapelier2024}. This case is particularly useful for the OpenMP target study because the mesh is regular enough to expose arithmetic throughput, while the high-order residual still exercises the same reconstruction, quadrature, Riemann solver, and explicit Runge--Kutta kernels used by production cases.

\begin{figure}[t]\begin{centering}

{\includegraphics[angle=0,width=0.99\columnwidth]{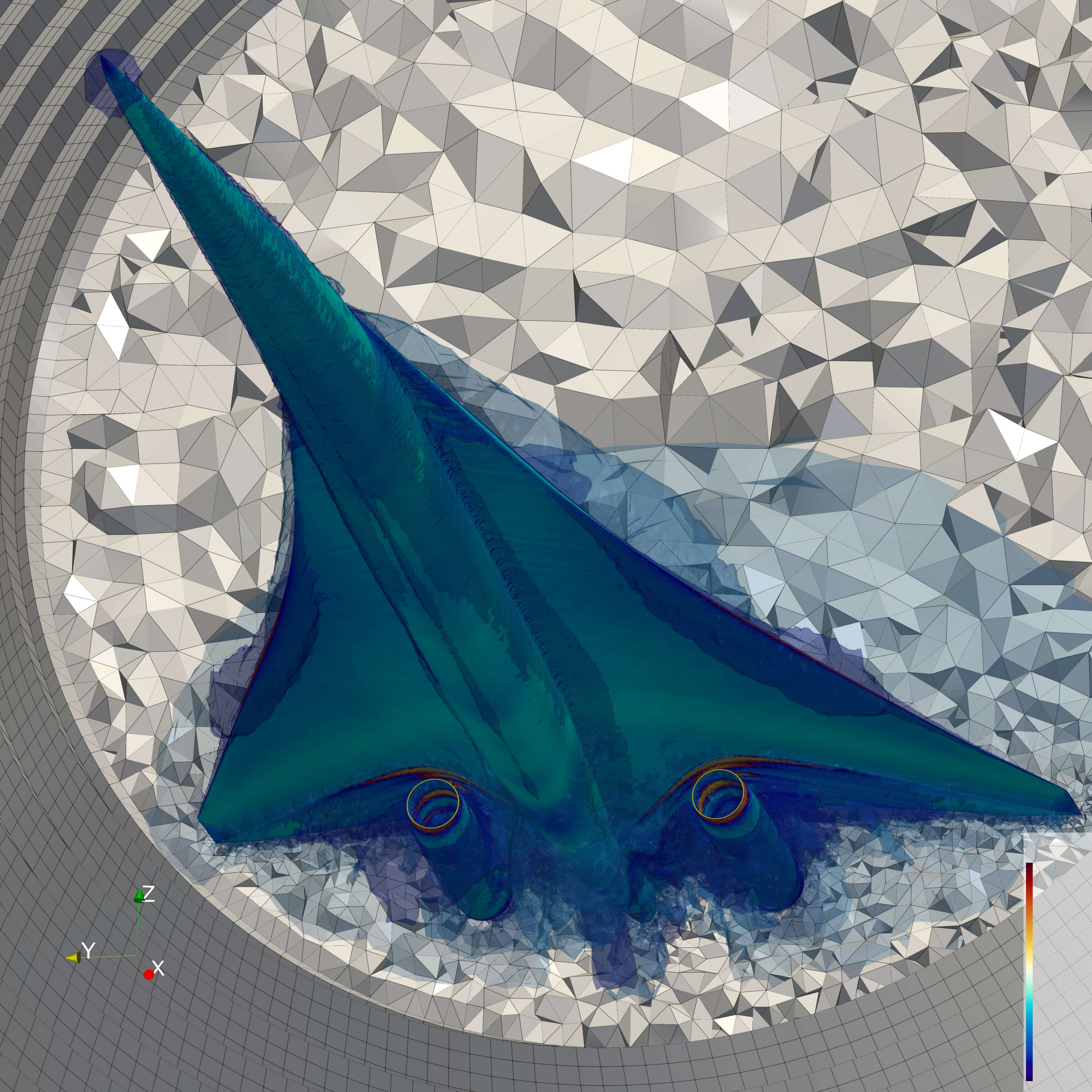}}\\
\par\end{centering}\caption{Isosurface of Mach number and surface contour plot of the pressure footprint for the LM1021 sonic-boom benchmark case.}\label{fig:boom}\end{figure}

\paragraph{\textbf{LM1021 supersonic sonic-boom configuration}}
The LM1021 case is based on the Lockheed Martin 1021-01 configuration from the first AIAA Sonic Boom Prediction Workshop, whose objective was to compare CFD predictions of the near-field pressure signature of a low-boom supersonic transport concept \citep{boom1}. The calculation uses the three-dimensional Euler equations at a freestream Mach number $M_\infty=1.6$ and model incidence of $2.3^\circ$. The domain uses supersonic inflow and outflow boundaries, wall boundaries on the aircraft surface, and a symmetry boundary condition because a half-model mesh is employed. The diagnostic of interest is the pressure signature $\Delta p/p_\infty$, including extraction along the centreline beneath the model, as in the workshop comparisons. In the present benchmark matrix the LM1021 case is represented by a 9 million-cell hybrid mesh and a $\mathcal{P}_3$ reconstruction, as shown in \Cref{fig:boom}, making it a useful intermediate step between canonical TGV turbulence and the much larger high-lift CRM meshes.

\paragraph{\textbf{NASA high-lift CRM}}
The NASA High-Lift Common Research Model (CRM-HL) is used to represent a large, realistic high-lift transport-aircraft calculation with deployed lifting surfaces and strongly three-dimensional separated flow. The AIAA CFD High-Lift Prediction Workshop series provides open geometries, committee grids, common test cases, and wind-tunnel data for assessing CFD methods for landing and take-off high-lift configurations \citep{nasaHLPW2026,hlpw5Website}. The CRM-HL geometry was defined as an open reference high-lift configuration for the CRM ecosystem \citep{LacyClark2020CRMHL}, with experimental databases including NASA Langley 14-by-22-foot testing of a 10\% semispan model \citep{Lin2022CRMHL} and the more recent 5.2\% CRM-HL datasets used in HLPW-5 and HLPW-6 \citep{nasaLavaCRMHL2026,hlpw6Cases2026}.

\begin{figure*}[t]\begin{centering}

{\includegraphics[angle=0,width=0.99\textwidth]{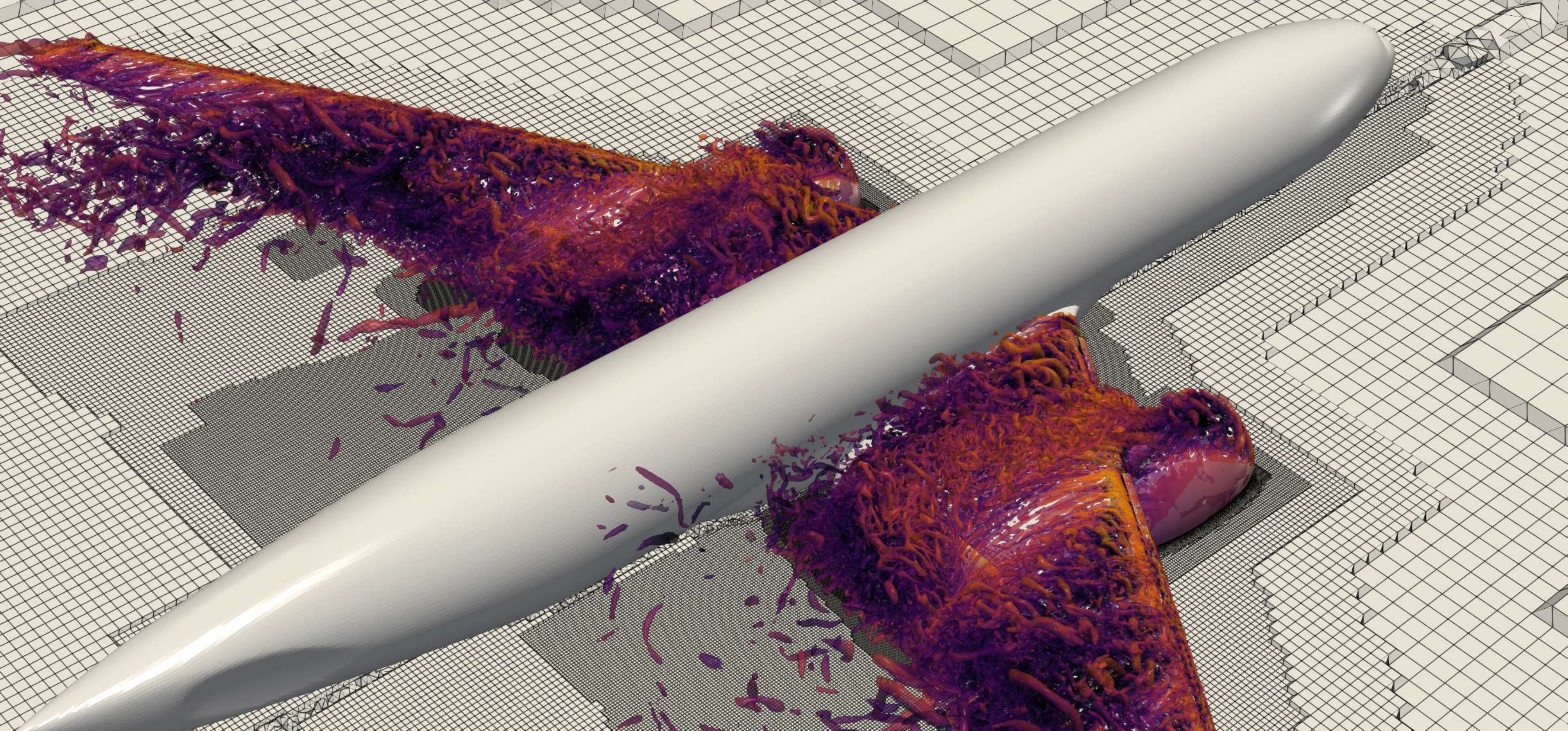}}
\par\end{centering}\caption{Isosurface of Q-criterion coloured by Mach number for the NASA High-Lift CRM configuration on the medium mesh used to assess the portability of the high-order implementation on production meshes.}\label{fig:crm_configuration}\end{figure*}

In the present work, the CRM-HL calculations use a tail-off landing configuration, with deployed slats and flaps and the nacelle/pylon/chine installation, but without the horizontal and vertical tail surfaces, as shown in \Cref{fig:crm_configuration}. This corresponds to the HLPW-5 tail-off landing-configuration family used for Reynolds-number studies. The NASA-CRM-C and NASA-CRM-M entries in \Cref{tab:benchmarks} use coarse and medium production-scale high-lift CRM hybrid meshes with $\mathcal{P}_4$ reconstruction. These cases therefore test whether the OpenMP target implementation can retain device-resident execution when the working set grows to billions of reconstructed degrees of freedom and billions of face quadrature-point flux evaluations per residual step.

\begin{table*}[t]
\centering
\caption{Benchmark matrix. Here M and B denote millions and billions, respectively. The face quadrature point count is also the number of Riemann problems per residual evaluation; a standard second-order finite-volume scheme solves approximately one Riemann problem per mesh face.}
\label{tab:benchmarks}
\scriptsize
\setlength{\tabcolsep}{3.0pt}
\renewcommand{\arraystretch}{1.12}
\begin{tabular*}{\textwidth}{@{\extracolsep{\fill}}lrrrr@{}}
\toprule
Case & Mesh & Mesh faces & Rec. DoF/var. & Face QPs \\
\midrule
TGV-$\mathcal{P}_2$        & 2M hexas     & 6M     & 54M     & 24M \\
TGV-$\mathcal{P}_3$        & 2M hexas     & 6M     & 74M     & 54M \\
TGV-$\mathcal{P}_4$        & 2M hexas     & 6M     & 176M    & 96M \\
TGV-$\mathcal{P}_5$        & 2M hexas     & 6M     & 210M    & 150M \\
LM1021-$\mathcal{P}_3$     & 9M hybrid    & 18M    & 279M    & 108M \\
NASA-CRM-C-$\mathcal{P}_4$ & 74M hybrid   & 222M   & 6.51B  & 3.55B \\
NASA-CRM-M-$\mathcal{P}_4$ & 370M hybrid  & 1.11B & 32.56B & 17.76B \\
\bottomrule
\end{tabular*}
\end{table*}

\subsection{Execution configuration}
\label{subsec:execution_configuration}

All performance results are reported as wall-clock time per solver iteration. Each benchmark timing is the mean of three independent runs of 100 explicit solver iterations. On \lumi, the run-to-run variation across these repeats was below 1\% for all reported benchmark timings. The LM1021 and NASA high-lift CRM cases are used here as production-scale performance benchmarks rather than as additional validation studies. Within the computational allocation available for this development study, these simulations were not advanced to the flow convergence required for meaningful comparison with experimental or reference results, although converged calculations for these configurations have been reported previously \citep{MALTSEV2023111755}. The present results are therefore used only to determine per-iteration cost and relative speed-up, and no aerodynamic quantities from these truncated runs are interpreted as converged results. On LUMI-G, the CPU-only runs use a fully populated node with the fastest measured hybrid MPI+OpenMP placement: eight MPI ranks per node and eight OpenMP threads per rank, using all 64 CPU cores of the node. This placement is used for both the \ucns\ v3 CPU reference and the \ucns\ v4 CPU build so that the CPU comparison reflects the effect of the data-layout and OpenMP restructuring rather than a change in node occupancy.

For the fully accelerated runs on LUMI-G, each node uses eight MPI ranks, with one MPI rank pinned to one AMD MI250X graphics compute die (GCD). Since a LUMI-G node exposes eight GCDs, this gives one rank per GCD and keeps the MPI decomposition aligned with the accelerator topology. The same global problem is used for each strong-scaling sequence, and the number of populated nodes is then increased. On the \archer\ GPU development platform, only a single-node accelerated $\mathcal{P}_2$ Taylor--Green vortex verification run is reported; it uses the corresponding one-rank-per-accelerator placement and serves as a development and portability check against the \lumi\ result rather than as part of the production performance dataset.

The profiling data report the percentage of the average iteration time spent in the principal solver components, including reconstruction, flux evaluation, update, halo exchange, and boundary treatment. These percentages are used to explain the scaling behaviour, while the primary performance metric remains the end-to-end wall-clock cost per complete explicit iteration.

For a run using $N$ nodes, the strong-scaling efficiency is defined as
\begin{equation}
  \eta_s(N) = \frac{T(N_0)}{T(N)}\frac{N_0}{N},
  \label{eq:strong_eff}
\end{equation}
where $T(N)$ is the wall-clock iteration time on $N$ nodes and $N_0$ is the baseline node count. The speed-up of a solver variant relative to a selected baseline is reported as
\begin{equation}
  S = \frac{T_{\mathrm{baseline}}}{T_{\mathrm{variant}}},
  \label{eq:speedup}
\end{equation}
with the baseline stated in each figure.

\section{Results}
\label{sec:results}

\subsection{Taylor--Green vortex reproducibility}
\label{subsec:verification}

\begin{figure*}[t]\begin{centering}

{\includegraphics[angle=0,width=0.49\textwidth]{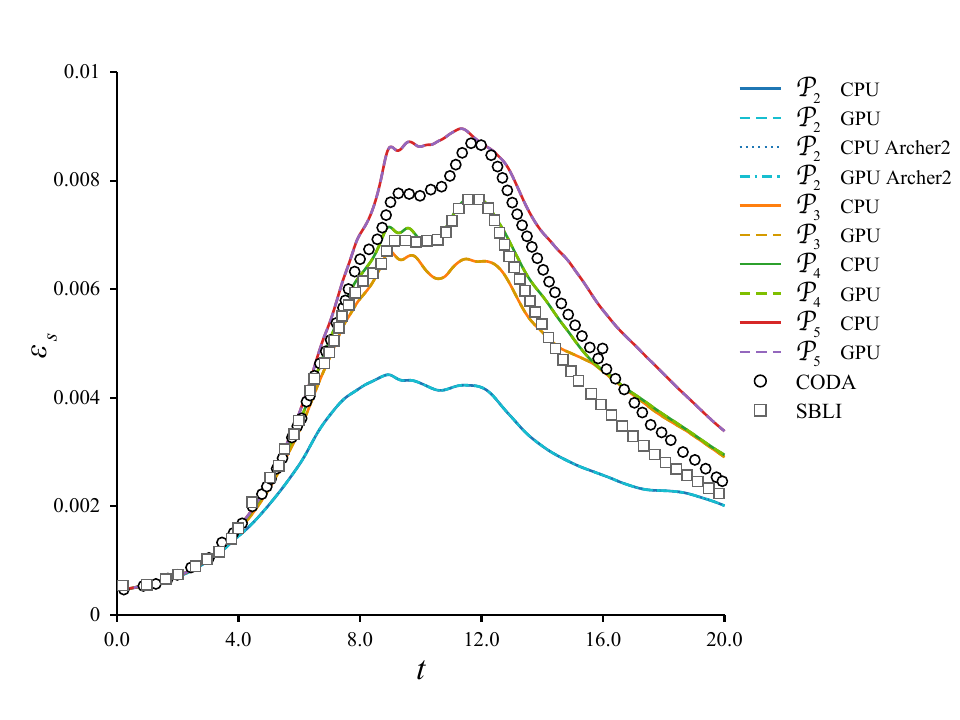}}
{\includegraphics[angle=0,width=0.49\textwidth]{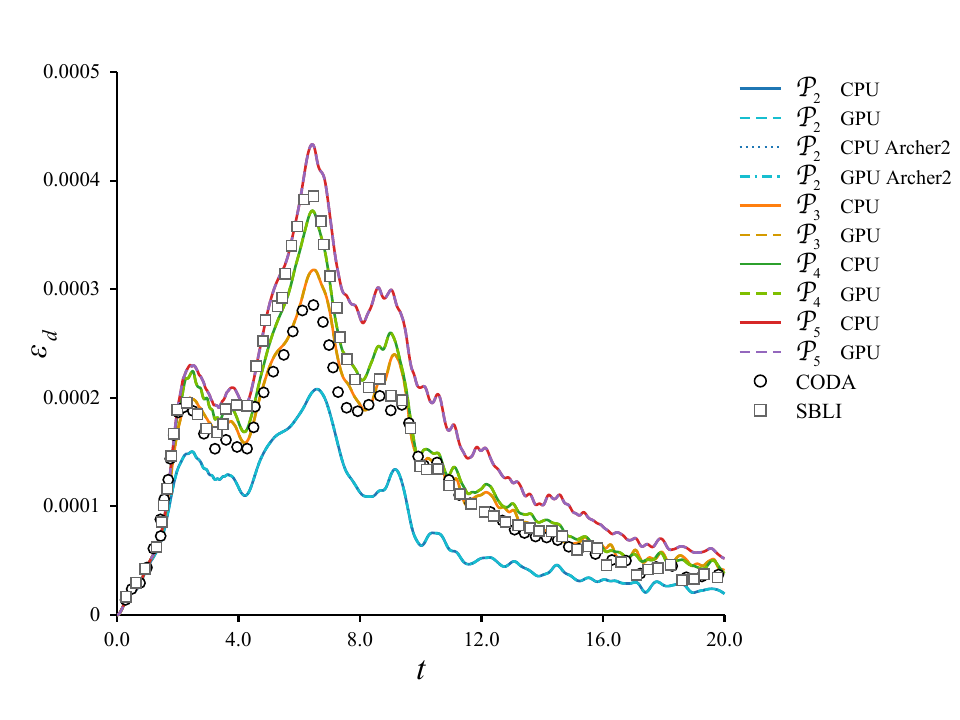}}
\par\end{centering}\caption{Solenoidal and dilatational dissipation histories for the Taylor--Green vortex benchmark using one node on ARCHER2 and two nodes on LUMI, and offers a good agreement as we increase the order with the modal DG-4th order scheme of the (CODA) CFD software and a 6th-order TENO scheme of the (OPENSBLI) CFD software as provided by Chapelier et al. \citep{Chapelier2024} at the same $(128^3)$ resolution and the results of Lusher and Sandham \cite{tgv_super}}\label{fig:tgv_dissipation_histories}\end{figure*}

The primary verification result is the comparison between accelerated and non-accelerated Taylor--Green vortex simulations in \Cref{fig:tgv_dissipation_histories}. The solenoidal and dilatational dissipation histories overlap for the corresponding CPU and GPU variants on \lumi, demonstrating that the OpenMP target path preserves the numerical behaviour of the CPU implementation. In addition, the \archer\ GPU development platform was used for a single-node $\mathcal{P}_2$ Taylor--Green vortex run, which reproduced the corresponding \lumi\ accelerated result to machine precision. This is a stronger check than agreement in a final scalar diagnostic, because the dissipation histories are sensitive to the reconstruction, viscous-gradient evaluation, flux integration, Runge--Kutta update, and halo exchange throughout the simulation.

The same figure also shows that the present results remain in close agreement  with the CODA and SBLI reference curves on $128^3$. The solenoidal dissipation peak and subsequent decay are captured consistently across reconstruction orders, while the dilatational dissipation remains much smaller and follows the expected compressible activity of the supersonic TGV\@. The performance results that follow can therefore be interpreted as measurements of equivalent solver variants, not as timings of a reduced or numerically altered GPU algorithm.

\subsection{Operation-level profiling on LUMI}
\label{subsec:lumi_profile}

\begin{figure*}[t]\begin{centering}

{\includegraphics[angle=0,width=0.33\textwidth]{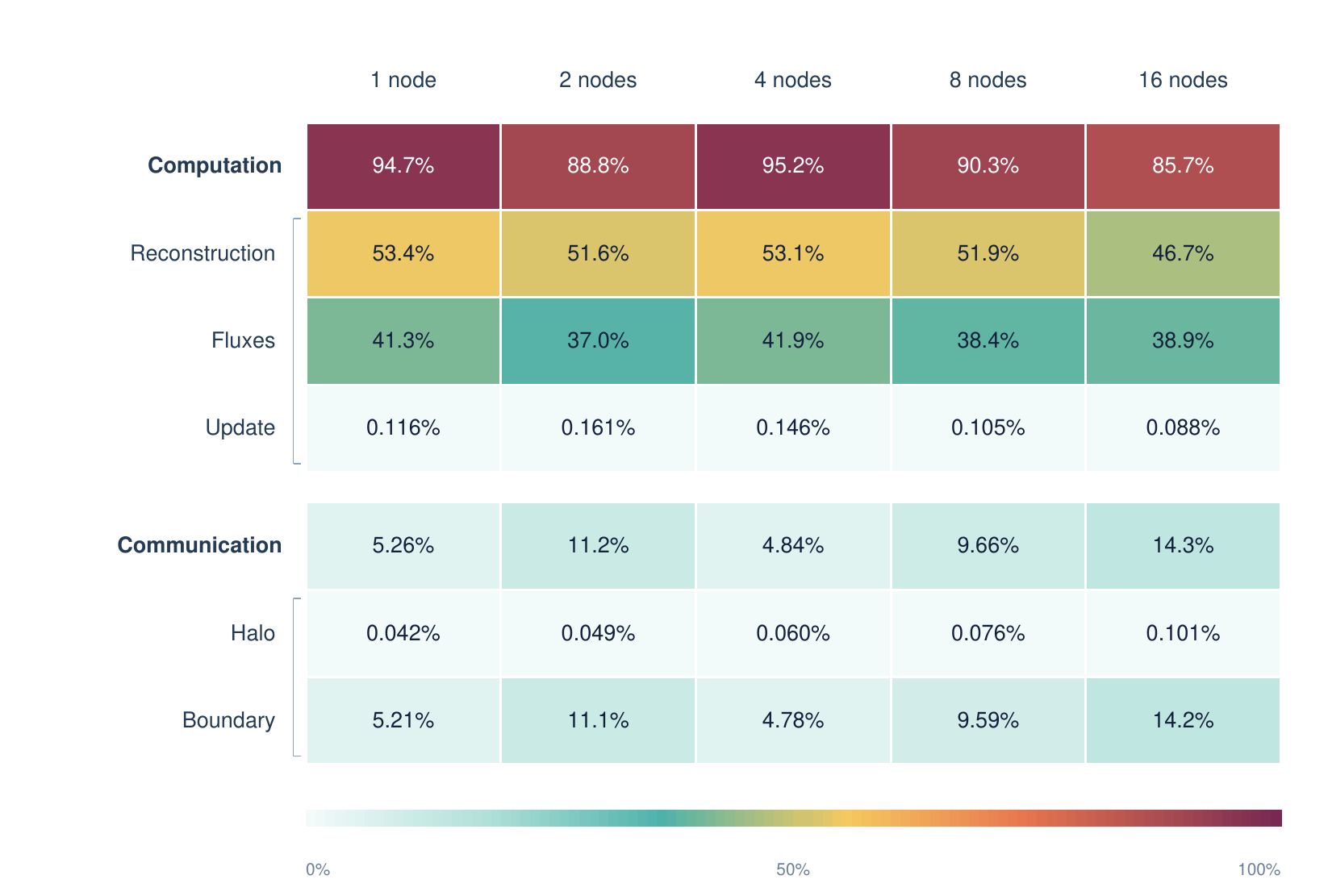}}
{\includegraphics[angle=0,width=0.33\textwidth]{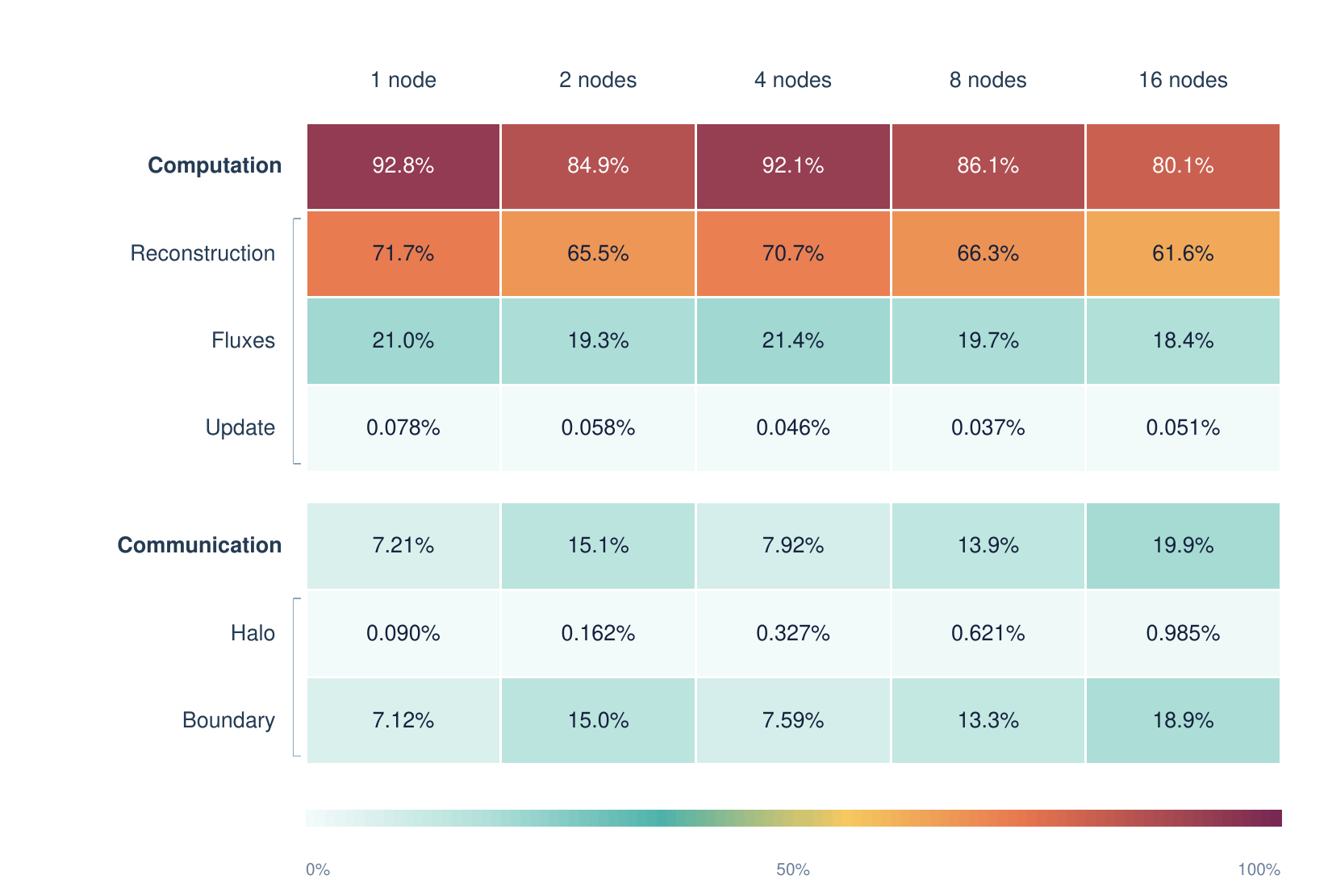}}
{\includegraphics[angle=0,width=0.33\textwidth]{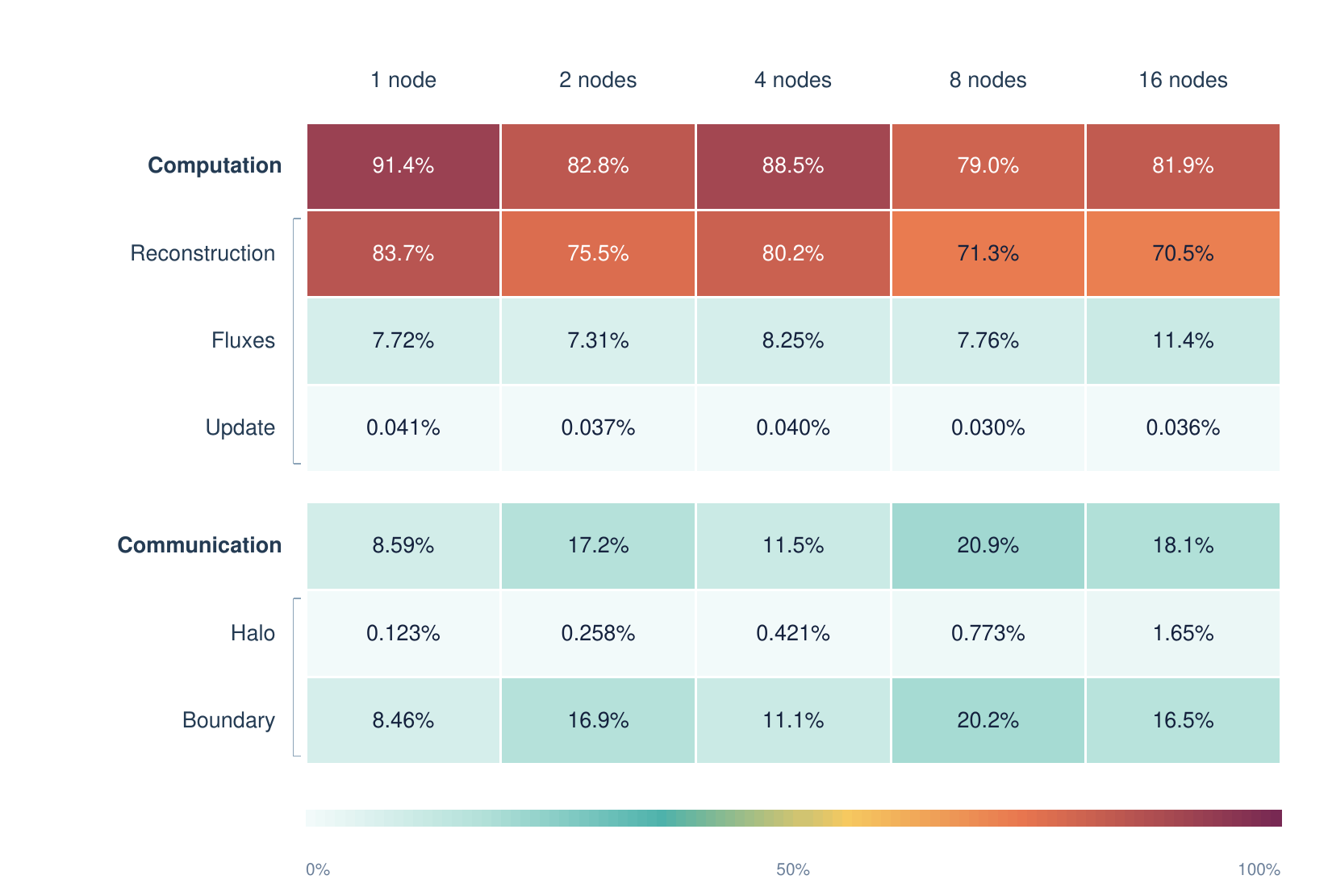}}
\par\end{centering}\caption{Operation-level runtime distribution for the baseline OpenMP v3 UCNS3D on CPUs (left), the OpenMP v4 UCNS3D on CPUs (middle), and the OpenMP target implementation in v4 UCNS3D on GPU nodes (right) for the Taylor-Green vortex flow problem using CWENOZ5 ($\mathcal{P}_{4}$).}\label{fig:operation_heatmap}\end{figure*}

\Cref{fig:operation_heatmap} reports the operation-level distribution of wall-clock iteration time for \ucns\ v3 CPU, \ucns\ v4 CPU, and \ucns\ v4 GPU runs on the same LUMI hardware over 1, 2, 4, 8, and 16 nodes for the Taylor-Green vortex flow problem using CWENOZ5 ($\mathcal{P}_{4}$). This comparison is important because it separates two effects of the port: the CPU-only consequences of the v4 restructuring and the additional acceleration obtained when the same v4 execution path is offloaded to GPUs.

For the v3 CPU baseline, computation accounts for 94.7\% of the iteration time on one node and 85.7\% on 16 nodes, with communication increasing from 5.26\% to 14.3\%. The dominant computational costs are reconstruction and flux evaluation: reconstruction decreases from 53.4\% to 46.7\%, while fluxes remain between 37.0\% and 41.9\%. In the v4 CPU build, the reconstructed-state machinery becomes a larger fraction of the total cost, accounting for 71.7\% on one node and 61.6\% on 16 nodes, while flux evaluation is reduced to about 18--21\%. This shift is consistent with the flat-array layout and loop restructuring: some of the original overheads are reduced, but the remaining high-order reconstruction work becomes more visible in the profile.

The v4 GPU profile shows the same trend more strongly. Reconstruction accounts for 83.7\% of the one-node iteration time and remains about 70.5\% at 16 nodes, while flux evaluation is reduced to 7.72\% on one node and 11.4\% at 16 nodes. The communication fraction increases from 8.59\% to values around 18--21\% as the node count grows. Most of this increase is associated with boundary and inter-partition work rather than the halo packing itself, whose measured share remains small. These profiles justify the scaling results: the GPU path accelerates the face-flux and update work effectively, but the end-to-end iteration time is then increasingly controlled by reconstruction and by the communication/boundary fraction exposed by strong scaling.

\subsection{GPU acceleration across the benchmark set}
\label{subsec:gpu_acceleration}

\begin{figure*}[t]\begin{centering}

{\includegraphics[angle=0,width=0.99\textwidth]{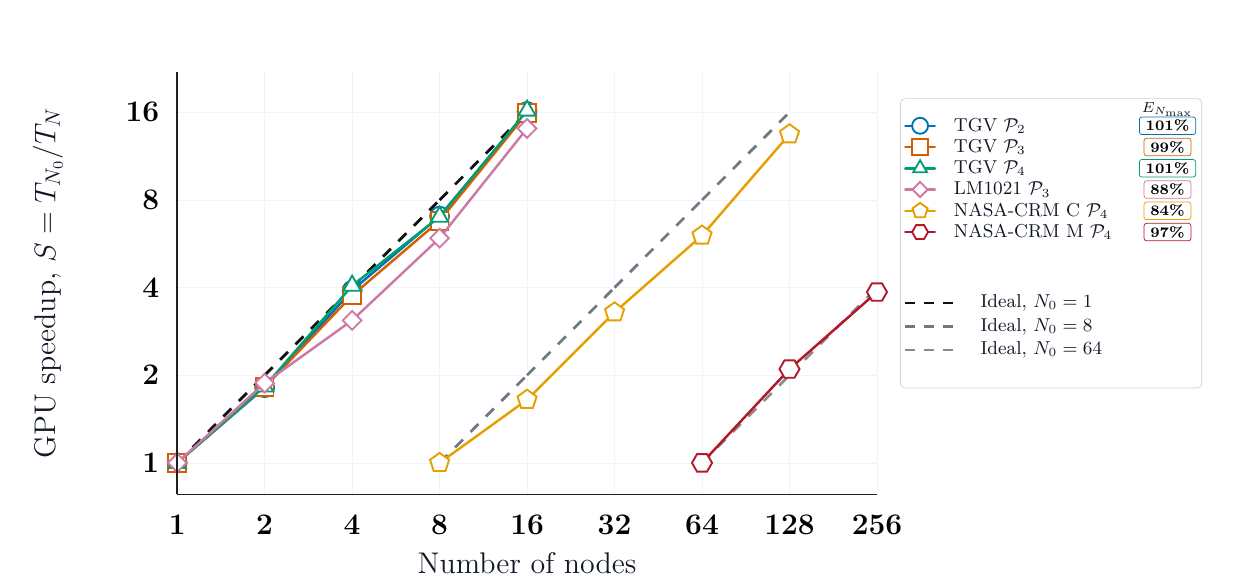}}
\par\end{centering}\caption{GPU speed-up summary across the benchmark set, highlighting the production impact of the OpenMP target port.}\label{fig:gpu_speedup_all_cases}\end{figure*}

\Cref{fig:gpu_speedup_all_cases} summarises the GPU strong-scaling behaviour for the full benchmark matrix. The TGV cases show near-ideal scaling at the largest tested node counts, with efficiencies of 101\%, 99\%, and 101\% for the $\mathcal{P}_2$, $\mathcal{P}_3$, and $\mathcal{P}_4$ reconstructions, respectively. These cases have regular hexahedral connectivity and a large amount of homogeneous reconstruction and flux work per device, so the communication-to-computation balance remains favourable.

The more complex aerodynamic benchmarks also retain useful efficiency. The LM1021-$\mathcal{P}_3$ case reaches 88\% efficiency, while the NASA-CRM-C-$\mathcal{P}_4$ and NASA-CRM-M-$\mathcal{P}_4$ cases reach 84\% and 97\%, respectively. The lower efficiency of the coarse CRM case is consistent with a smaller amount of work per device and a stronger relative contribution from hybrid-mesh boundary and partition interfaces. The medium CRM case recovers higher efficiency because the larger production mesh provides more local reconstruction and flux work per GCD\@. Overall, the benchmark-wide result shows that the OpenMP target implementation scales beyond the canonical TGV case and remains effective for hybrid unstructured production meshes.

\subsection{Effect of the v4 restructuring on CPU and GPU performance}
\label{subsec:tgv_scaling_results}

\begin{figure*}[t]\begin{centering}

{\includegraphics[angle=0,width=0.99\textwidth]{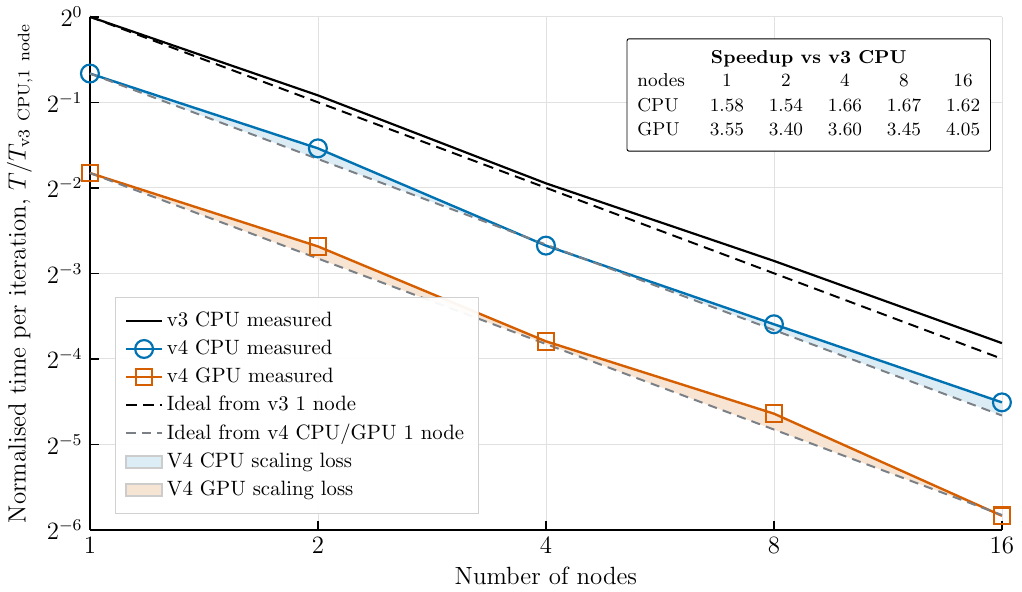}}
\par\end{centering}\caption{Strong-scaling speed-up for the Taylor--Green vortex benchmark using the $\mathcal{P}_3$ reconstruction.}\label{fig:tgv_p3_speedup}\end{figure*}

\begin{figure*}[t]\begin{centering}

{\includegraphics[angle=0,width=0.99\textwidth]{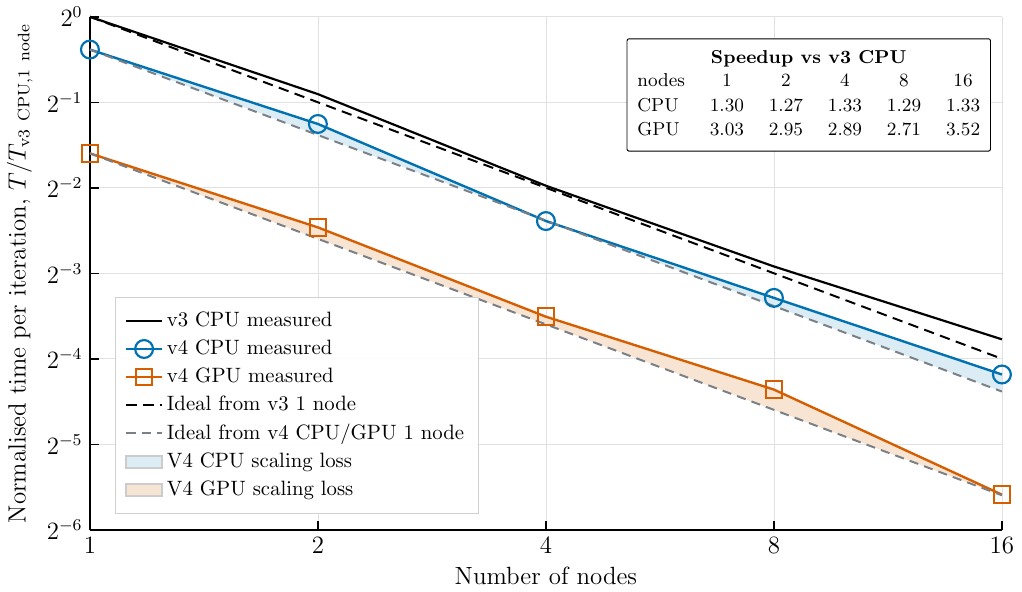}}
\par\end{centering}\caption{Strong-scaling speed-up for the Taylor--Green vortex benchmark using the $\mathcal{P}_4$ reconstruction.}\label{fig:tgv_p4_speedup}\end{figure*}

\Cref{fig:tgv_p3_speedup,fig:tgv_p4_speedup} are the central performance results because they show that the v4 changes improve the CPU-only solver as well as enabling GPU acceleration. For the $\mathcal{P}_3$ TGV case, the v4 CPU build is consistently faster than the v3 CPU baseline, with same-node speed-ups of about 1.54--1.67 across the tested range. The v4 GPU build increases the same-node speed-up to about 3.40--4.05, with the largest value observed at 16 nodes. Thus, the GPU gain is added on top of a substantial CPU-side improvement from the flat run-time layout, revised loop structure, and reduced memory footprint.

The $\mathcal{P}_4$ case shows the same qualitative behaviour, but with smaller CPU-side gains. The v4 CPU build gives speed-ups of about 1.27--1.33 over v3 CPU, while the v4 GPU build gives about 2.71--3.52. This reduction relative to $\mathcal{P}_3$ is consistent with the heavier reconstruction workload, larger temporary storage requirements, and increased memory traffic at higher order. Even so, the v4 CPU implementation remains faster than the original CPU code at every node count, and the GPU implementation maintains a clear end-to-end advantage. These figures therefore show that the port should not be interpreted as a GPU-only exercise: the software restructuring improves the conventional CPU execution path and then exposes additional acceleration on the GPUs.

\section{Discussion}
\label{sec:discussion}

\subsection{Portability versus peak performance}
\label{subsec:portability}

The results show that the OpenMP target port is best understood as a full run-time modernisation rather than as a collection of isolated GPU directives. The exact agreement of the accelerated and non-accelerated TGV results in \Cref{fig:tgv_dissipation_histories} demonstrates that the numerical path has been preserved, while \Cref{fig:tgv_p3_speedup,fig:tgv_p4_speedup} show that the same changes also improve the CPU-only implementation. This is important for a production Fortran solver: if the GPU branch required a separate numerical path, the cost of maintaining and validating the code would grow rapidly.

The achieved speed-ups do not imply that directive-based OpenMP automatically matches a hand-tuned architecture-specific implementation. They show instead that a standards-based, single-source strategy can deliver useful end-to-end acceleration while preserving the CPU route. The fixed LUMI-G placements used here make this comparison deliberately strict: the CPU runs use all 64 CPU cores through the best measured 8 MPI rank by 8 thread layout, and the GPU runs use all eight MI250X GCDs through one MPI rank per GCD\@. The performance difference is therefore a comparison between fully populated node configurations.

\subsection{Bottlenecks and lessons learned}
\label{subsec:lessons}

The operation profiles in \Cref{fig:operation_heatmap} explain both the speed-ups and the remaining limits. On the GPU, flux evaluation becomes a comparatively small part of the iteration, while reconstruction becomes the dominant cost. This is expected for high-order finite-volume schemes on unstructured meshes: reconstruction contains indirect stencil access, local polynomial operations, nonlinear weights, and order-dependent temporary storage. These features give more arithmetic work than a second-order method, but they do not turn the solver into a dense linear-algebra workload. Memory bandwidth, memory latency, and irregular access patterns therefore remain central limitation of the unstructured mixed-element meshes used. 

The second limiting factor is the strong-scaling communication fraction. As the problem is divided over more nodes, the local work per rank and per GCD decreases, and boundary and inter-partition operations occupy a larger share of each iteration. The halo-packing cost itself remains small in the profiles, which supports the decision to use GPU-aware MPI and device-resident buffers. The local maximum in the communication fraction at two nodes is consistent with the first introduction of inter-node transfers on LUMI, amplified by the partition interfaces and rank-to-GCD mapping at that scale. Its reduction at four nodes reflects a more favourable communication-to-computation balance rather than necessarily a lower absolute communication cost; because the feature recurs across all implementations and three independent 100-iteration averages, it is more plausibly a systematic decomposition/topology effect than a cache or sampling artefact. However, the boundary and communication share still grows with node count, so further gains at scale will require attention to partition quality, overlap of communication and computation, and reduction of boundary-side work.

The comparison between $\mathcal{P}_3$ and $\mathcal{P}_4$  as shown in \Cref{fig:tgv_p3_speedup} and \Cref{fig:tgv_p4_speedup} respectively, also shows that higher polynomial order does not guarantee proportionally larger GPU speed-up. The $\mathcal{P}_4$ case carries more reconstruction work and larger private temporaries, which increases register and memory pressure as well as indirect memory access. The port therefore benefits from case-specialised compile-time bounds and from keeping local algebra explicit, but reconstruction remains the kernel family where further optimisation will have the largest effect.

It must be stressed that the observed 2.71–4.05× node-level speed-up should not be compared directly with the approximately 60–70× ratio of theoretical GPU-to-CPU hardware peaks. Peak FP64 throughput assumes sustained, regular vector or matrix operations, while peak HBM bandwidth assumes contiguous and coalesced streaming access; neither condition characterises the dominant UCNS3D kernels. Reconstruction and flux evaluation involve indirect stencil gathers, irregular and limited reuse, small order-dependent local operations, conditional control flow, and private temporaries that increase register pressure and can reduce occupancy. The end-to-end measurements also include boundary treatment, synchronisation, and MPI communication, whose costs do not decrease in proportion to GPU FLOPS or HBM bandwidth. The measured acceleration therefore represents a substantial application-level improvement over a fully populated 64-core CPU baseline for a production CFD workload, rather than a claim of approaching the accelerator’s dense-arithmetic peak.

\subsection{Implications for future exascale systems}
\label{subsec:future}

Future exascale systems are likely to be more heterogeneous, not less. For high-order CFD solvers, this means that performance portability must become part of the numerical software design rather than an afterthought. The present results suggest that OpenMP can provide a practical route when the solver is refactored around explicit data lifetime, flat run-time arrays, regular outer loops, target-callable routines, and GPU-aware MPI communication.

The benchmark-wide results in \Cref{fig:gpu_speedup_all_cases} are particularly promising because they cover both compact structured TGV cases and substantially larger hybrid unstructured aerodynamic meshes. Although the production cases naturally exhibit different scaling behaviour from the canonical case, they maintain strong practical efficiency and reveal a consistent set of optimisation opportunities. This suggests that the current OpenMP target strategy provides a robust foundation, with further gains expected from focused improvements to reconstruction, boundary processing, communication overlap, and rank/GCD placement.

The 2.71–4.05× same-node acceleration is substantial in this context because it is measured end-to-end against all 64 CPU cores and includes reconstruction, flux integration, boundary treatment, and MPI communication. The remaining gap relative to theoretical hardware peak ratios is consistent with the indirect addressing, irregular memory access, small local operations, branch divergence, and register pressure inherent to high-order unstructured CFD. These results therefore demonstrate practical and portable full-solver acceleration while identifying reconstruction locality and temporary-storage pressure as the principal targets for further optimisation.

These results also point to a clear path for future development. Adaptive algorithms, implicit solvers, dynamic load balancing, and mixed-precision strategies can now be pursued within the same accelerator-ready framework, with careful attention to their numerical and performance implications. The one-language, one-directive-model approach therefore offers a sustainable route for extending UCNS3D on heterogeneous systems while keeping the scientific algorithm visible in Fortran and avoiding a permanent split between CPU and GPU solver branches.

\section{Conclusions}
\label{sec:conclusions}

This paper has presented the transition of \ucns\ from the v3 OpenMP CPU implementation to the v4 OpenMP target GPU implementation under the principle of one language, one directive model, and one code base. The technical strategy combines MPI domain decomposition with OpenMP threading for CPU builds and OpenMP target offload for the \lumi\ GPU runs, while the \archer\ GPU development platform provides an independent single-node $\mathcal{P}_2$ portability check. The main implementation themes are persistent device data management, flat run-time arrays, compile-time sizing of temporary storage, GPU-aware MPI halo exchange, target-compatible Fortran call structures, explicit target-clause auditing, and acceleration of the complete explicit reconstruction, gradient, flux, boundary, communication, and update path.

The port also audited OpenMP accelerated-region syntax and target data clauses, including variable scope, mapping, and declarations within device regions. Where appropriate, automated code-review assistants supported source navigation and checklist-based inspection, while all findings remained subject to developer-led verification.

The results demonstrate that the accelerated solver preserves the numerical behaviour of the CPU implementation while reducing wall-clock iteration time on current heterogeneous systems. The Taylor--Green vortex dissipation histories are reproducible between accelerated and non-accelerated variants to machine precision on \lumi, and the single-node $\mathcal{P}_2$ \archer\ GPU development run independently reproduces the corresponding \lumi\ result. The strong-scaling measurements show that the v4 restructuring improves the CPU-only code before additional GPU acceleration is applied. More broadly, the work shows that non-vendor-specific OpenMP can be a viable path for preparing mature high-order CFD solvers for future exascale architectures, provided that the port is treated as a full run-time transformation rather than a kernel-by-kernel annotation exercise. For \ucns, the decisive changes were keeping data resident on the device, replacing GPU-unfriendly derived-type access with flat arrays, making temporary storage explicit at compilation, avoiding non-portable dependencies for small local algebra, and communicating halo data directly from device buffers. Building on this foundation, future work will pursue accelerating the implicit time stepping, real-gas models, and introducing mixed-precision arithmetic within the same single-source OpenMP framework.

\section*{Acknowledgements}

Software development and associated technical staff effort underpinning this work were funded under the embedded Computational Science and Engineering (eCSE) programme of the UKRI-funded ARCHER2 UK National Supercomputing Service, through project ARCHER2-eCSE11-3, ``Evolving UCNS3D CFD Software to Heterogeneous Exascale Systems via OpenMP''. This work used the ARCHER2 UK National Supercomputing Service (https://www.archer2.ac.uk). We also acknowledge the EuroHPC Joint Undertaking for awarding project ID EHPC-DEV-2025D04-111, ``High-Order Finite-Volume Optimisation'' (HOVE4), access to the EuroHPC supercomputer LUMI, hosted by CSC (Finland) and the LUMI consortium, through a EuroHPC Development Access call.

\section*{Data availability}

The datasets, meshes, input files, and post-processing scripts used in this study will be made available at a free-access repository upon publication.

\bibliographystyle{elsarticle-num}
\bibliography{all_references}

@article{ANTONIADIS2022108453,
title = {{UCNS3D: An open-source high-order finite-volume unstructured CFD solver}},
journal = {Computer Physics Communications},
volume = {279},
pages = {108453},
year = {2022},
issn = {0010-4655},
doi = {https://doi.org/10.1016/j.cpc.2022.108453},
author = {Antonis F. Antoniadis and Dimitris Drikakis and Pericles S. Farmakis and Lin Fu and Ioannis Kokkinakis and Xesús Nogueira and Paulo A.S.F. Silva and Martin Skote and Vladimir Titarev and Panagiotis Tsoutsanis},
}

@ARTICLE{boom1,
author={Park, M.A. and Morgenstern, J.M.},
title={Summary and statistical analysis of the first {AIAA} sonic boom prediction workshop},
journal={Journal of Aircraft},
year={2016},
volume={53},
number={2},
pages={578-598},
doi={10.2514/1.C033449},
}

@article{BORGES20083191,
title = {An improved weighted essentially non-oscillatory scheme for hyperbolic conservation laws},
journal = {Journal of Computational Physics},
volume = {227},
number = {6},
pages = {3191-3211},
year = {2008},
issn = {0021-9991},
doi = {https://doi.org/10.1016/j.jcp.2007.11.038},
author = {Rafael Borges and Monique Carmona and Bruno Costa and Wai Sun Don},
}

@ARTICLE{Chapelier2024,
	author = {Chapelier, Jean-Baptiste and Lusher, David J. and Van Noordt, William and Wenzel, Christoph and Gibis, Tobias and Mossier, Pascal and Beck, Andrea and Lodato, Guido and Brehm, Christoph and Ruggeri, Matteo and Scalo, Carlo and Sandham, Neil},
	title = {Comparison of high-order numerical methodologies for the simulation of the supersonic Taylor-Green vortex flow},
	year = {2024},
	journal = {Physics of Fluids},
	volume = {36},
	number = {5},
	doi = {10.1063/5.0206359},
	type = {Article},
}

@ARTICLE{DG_FV_MULTI,
	author = {Maltsev, Vadim and Skote, Martin and Tsoutsanis, Panagiotis},
	title = {High-order hybrid {D}{G}-FV framework for compressible multi-fluid problems on unstructured meshes},
	year = {2024},
	journal = {Journal of Computational Physics},
	volume = {502},
	pages = {},
	doi = {10.1016/j.jcp.2024.112819},
	type = {Article},
}

@ARTICLE{Dumbser2007204,
author={Dumbser, M. and Kaser, M. and Titarev, V.A. and Toro, E.F.},
title={Quadrature-free non-oscillatory finite volume schemes on unstructured meshes for nonlinear hyperbolic systems},
journal={Journal of Computational Physics},
year={2007},
volume={226},
number={1},
pages={204-243},
OPTnote={cited By (since 1996) 40},
OPTurl={http://www.scopus.com/inward/record.OPTurl?eid=2-s2.0-34548397128&partnerID=40&md5=624634fa8125b07237afc97638a1b4dd},
}

@Article{fluids9020033,
AUTHOR = {Adebayo, Ebenezer Mayowa and Tsoutsanis, Panagiotis and Jenkins, Karl W.},
TITLE = {Application of Central-Weighted Essentially Non-Oscillatory Finite-Volume Interface-Capturing Schemes for Modeling Cavitation Induced by an Underwater Explosion},
JOURNAL = {Fluids},
VOLUME = {9},
YEAR = {2024},
NUMBER = {2},
ARTICLE-NUMBER = {33},
ISSN = {2311-5521},
DOI = {10.3390/fluids9020033}
}

@ARTICLE{Gottlieb199873,
author={Gottlieb, S. and Shu, C.-W.},
title={Total variation diminishing {R}unge-{K}utta schemes},
journal={Mathematics of Computation},
year={1998},
volume={67},
number={221},
pages={73-85},
doi={10.1090/S0025-5718-98-00913-2},
}

@ARTICLE{HLLC,
author={Toro, E.F. and Spruce, M. and Speares, W.},
title={Restoration of the contact surface in the {H}{L}{L}-{R}iemann solver},
journal={Shock Waves},
year={1994},
volume={4},
number={1},
pages={25-34},
}

@misc{hlpw5Website,
  author = {{AIAA CFD High Lift Prediction Workshop Committee}},
  title = {{5th AIAA CFD High Lift Prediction Workshop (HLPW-5)}},
  year = {2024},
  url = {https://hiliftpw.larc.nasa.gov/index.html},
  note = {Accessed July 22, 2026}
}

@misc{hlpw6Cases2026,
  author = {{AIAA CFD High Lift Prediction Workshop Committee}},
  title = {{HLPW6 Test Cases}},
  year = {2026},
  url = {https://aiaa-hlpw.org/HLPW6/cases},
  note = {Last updated July 1, 2026; accessed July 22, 2026}
}

@inproceedings{LacyClark2020CRMHL,
  author = {Lacy, Doug S. and Clark, Adam M.},
  title = {Definition of Initial Landing and Takeoff Reference Configurations for the High Lift Common Research Model ({CRM-HL})},
  booktitle = {AIAA AVIATION 2020 Forum},
  year = {2020},
  doi = {10.2514/6.2020-2771}
}

@article{LevyPuppoRusso1999, 
title={Central WENO schemes for hyperbolic systems of conservation laws}, 
volume={33},
DOI={10.1051/m2an:1999152},
number={3},
journal={ESAIM: Mathematical Modelling and Numerical Analysis},
author={Levy, Doron and Puppo, Gabriella and Russo, Giovanni},
year={1999},
pages={547-571}
}

@techreport{Lin2022CRMHL,
  author = {Lin, John C. and Melton, LaTunia P. and Hannon, Judith A. and Koklu, Mehti and Andino, Marlyn Y.},
  title = {Semispan Test Results of a Conventional High-Lift Common Research Model in Landing Configuration},
  institution = {NASA Langley Research Center},
  year = {2022},
  number = {NASA/TP-20220008270},
  url = {https://ntrs.nasa.gov/citations/20220008270}
}

@article{MALTSEV2023111755,
author = {Vadim Maltsev and Dean Yuan and Karl W. Jenkins and Martin Skote and Panagiotis Tsoutsanis},
title = {Hybrid discontinuous {G}alerkin-finite volume techniques for compressible flows on unstructured meshes},
journal = {Journal of Computational Physics},
year = {2023},
volume = {473},
pages = {111755},
doi = {https://doi.org/10.1016/j.jcp.2022.111755},
}

@misc{MPI,
  title = {{M}{P}{I}},
  howpublished = {\url{https://www.mpi-forum.org/docs/mpi-4.0/mpi40-report.pdf}},
  note = {Accessed: 2021-10-10}
}

@misc{nasaHLPW2026,
  author = {{NASA}},
  title = {{AIAA CFD High-Lift Prediction Workshop}},
  year = {2026},
  url = {https://www.nasa.gov/reference/aiaa-cfd-high-lift-prediction-workshop/},
  note = {Page last updated January 30, 2026; accessed July 22, 2026}
}

@misc{nasaLavaCRMHL2026,
  author = {{NASA Advanced Supercomputing Division}},
  title = {High-Lift Common Research Model ({CRM-HL}) {WMLES}},
  year = {2026},
  url = {https://www.nas.nasa.gov/LAVA/unst_docs/tutorials/crm_highlift/crm_highlift/},
  note = {Page last updated June 15, 2026; accessed July 22, 2026}
}

@misc{openmp52,
  author = {{OpenMP Architecture Review Board}},
  title = {{OpenMP Application Programming Interface Version 5.2}},
  year = {2021},
  url = {https://www.openmp.org/specifications/}
}

@ARTICLE{pyfr,
author={Witherden, F.D. and Farrington, A.M. and Vincent, P.E.},
title={PyFR: An open source framework for solving advection-diffusion type problems on streaming architectures using the flux reconstruction approach},
journal={Computer Physics Communications},
year={2014},
volume={185},
number={11},
pages={3028-3040},
doi={10.1016/j.cpc.2014.07.011},
}

@article{Moxey2020Nektar,
author = {Moxey, David and Cantwell, Chris D. and Bao, Yan and Cassinelli, Andrea and Castiglioni, Giacomo and Chun, Eun-Jin and Juda, Emmanuel and Kazemi, Ehsan and Lackhove, Kristoffer and Marcon, Jonathan and Mengaldo, Gianmarco and Serson, Douglas and Turner, Michael and Xu, Haohua and Peir{\'o}, Joaquim and Kirby, Robert M. and Sherwin, Spencer J.},
title = {{Nektar++}: Enhancing the capability and application of high-fidelity spectral/$hp$ element methods},
journal = {Computer Physics Communications},
volume = {249},
pages = {107110},
year = {2020},
doi = {10.1016/j.cpc.2019.107110},
}

@misc{Renner2026NektarPerfPortable,
author = {Renner, Diego and Xing, Jacques and Xia, Boyang and W{\"u}stenberg, Henrik and Ye, Junjie and Moxey, David and Kirby, Mike and Sherwin, Spencer and Cantwell, Chris},
title = {From Kernels to Solvers: Performance-Portable High-Order PDE Simulation in {Nektar++}},
howpublished = {Poster, ARCHER2 Celebration of Science},
year = {2026},
publisher = {Zenodo},
doi = {10.5281/zenodo.19254088},
}

@article{Ferrer2023HORSES3D,
author = {Ferrer, Esteban and Rubio, Gonzalo and Ntoukas, Georgios and Laskowski, Wiktor and Mari{\~n}o, Oscar A. and Colombo, Santiago and Mateo-Gab{\'i}n, Andr{\'e}s and Marbona, Hugo and Manrique de Lara, Fernando and Huergo, Daniel and Manzanero, Javier and Rueda-Ram{\'i}rez, Andr{\'e}s M. and Kopriva, David A. and Valero, Eusebio},
title = {{HORSES3D}: A high-order discontinuous Galerkin solver for flow simulations and multi-physics applications},
journal = {Computer Physics Communications},
volume = {287},
pages = {108700},
year = {2023},
doi = {10.1016/j.cpc.2023.108700},
}

@misc{horses3dGpu,
author = {{NUMATH Group}},
title = {{HORSES3D GPU repository}},
year = {2026},
howpublished = {\url{https://github.com/loganoz/horses3d-gpu/}},
note = {Accessed 28 July 2026},
}

@article{Kurz2025GALAEXI,
author = {Kurz, Marius and Kempf, Dominik and Blind, Michael P. and Kopper, Patrick and Offenh{\"a}user, Philipp and Schwarz, Anna and Starr, Spencer and Keim, Jens and Beck, Andrea},
title = {{GAL{\AE}XI}: Solving complex compressible flows with high-order discontinuous Galerkin methods on accelerator-based systems},
journal = {Computer Physics Communications},
volume = {306},
pages = {109388},
year = {2025},
doi = {10.1016/j.cpc.2024.109388},
}

@article{RICCI2020105648,
title = {Hovering rotor solutions by high-order methods on unstructured grids},
journal = {Aerospace Science and Technology},
volume = {97},
pages = {105648},
year = {2020},
issn = {1270-9638},
doi = {https://doi.org/10.1016/j.ast.2019.105648},
author = {Francesco Ricci and Paulo A.S.F. Silva and Panagiotis Tsoutsanis and Antonis F. Antoniadis},
}

@article{SILVA2021106518,
title = {Simple multiple reference frame for high-order solution of hovering rotors with and without ground effect},
journal = {Aerospace Science and Technology},
volume = {111},
pages = {106518},
year = {2021},
issn = {1270-9638},
doi = {https://doi.org/10.1016/j.ast.2021.106518},
author = {Paulo A.S.F. Silva and Panagiotis Tsoutsanis and Antonis F. Antoniadis},
}

@ARTICLE{Silva2022,
author={Silva,P. A.S.F.  and Tsoutsanis, P.  and Antoniadis, A.F},
title={Numerical investigation of full helicopter with and without ground effect},
journal={Aerospace Science and Technology},
year={2022},
volume={122},
number={2},
doi={10.1016/j.ast.2022.107401},
art_number={107401},
}

@ARTICLE{Silva2024,
	author = {Silva, Paulo Augusto Strobel Freitas Da and Tsoutsanis, Panagiotis and Pinheiro Vaz, Jerson Rogério and Macias, Marianela M.},
	title = {A comprehensive CFD investigation of tip vortex trajectory in shrouded wind turbines using compressible {R}{A}{N}{S} solver},
	year = {2024},
	journal = {Energy},
	volume = {294},
	pages = {},
	doi = {10.1016/j.energy.2024.130929},
	type = {Article}
}

@ARTICLE{SSP4,
author={Spiteri, R. J. and Ruuth, S. J.},
title={A new class of optimal high-order strong-stability-preserving time-stepping schemes},
journal={SIAM Journal of Numerical Analysis},
year={2002},
volume={40},
number={2},
pages={469-491},
}

@ARTICLE{tgv_super,
author={Lusher, D.J. and Sandham, N.D.},
title={Assessment of low-dissipative shock-capturing schemes for the compressible {T}aylor-{G}reen vortex},
journal={AIAA Journal},
year={2021},
volume={59},
number={2},
pages={533-545},
doi={10.2514/1.J059672},
}

@ARTICLE{Titarev_cicp,
author={Titarev, V.A. and Tsoutsanis, P. and Drikakis, D.},
title={{W}{E}{N}{O} schemes for mixed-element unstructured meshes},
journal={Communications in Computational Physics},
year={2010},
volume={8},
number={3},
pages={585-609},
}

@CONFERENCE{Tommaso2024dynamic,
  title={Dynamic swirl distortion characteristics in {S}-shaped diffusers using {UCNS3D} and time-resolved, stereo {PIV} methods},
  author={Tommaso Piovesan and Matteo Migliorini and Pavlos K. Zachos and Panagiotis Tsoutsanis},
  booktitle={AIAA SCITECH 2024 Forum},
  year={2024},
  pages = {5-9},
}

@article{TSOUTSANIS2015207,
title = {Comparison of structured- and unstructured-grid, compressible and incompressible methods using the vortex pairing problem},
journal = {Computer Methods in Applied Mechanics and Engineering},
volume = {293},
pages = {207-231},
year = {2015},
issn = {0045-7825},
doi = {https://doi.org/10.1016/j.cma.2015.04.010},
author = {Panagiotis Tsoutsanis and Ioannis W. Kokkinakis and László Könözsy and Dimitris Drikakis and Robin J.R. Williams and David L. Youngs},
}

@article{TSOUTSANIS201869,
author = {Panagiotis Tsoutsanis},
title = {Extended bounds limiter for high-order finite-volume schemes on unstructured meshes},
journal = {Journal of Computational Physics},
year = {2018},
volume = {362},
pages = {69-94},
doi = {https://doi.org/10.1016/j.jcp.2018.02.009},
}

@article{TSOUTSANIS20218964,
title = {{CWENO Finite-Volume Interface Capturing Schemes for Multicomponent Flows Using Unstructured Meshes}},
journal = {Journal of Scientific Computing},
volume = {89},
pages = {64},
year = {2021},
doi = {https://doi.org/10.1007/s10915-021-01673-y},
author = {Panagiotis Tsoutsanis and Adebayo Ebenezer Mayowa and Merino Adrian Carriba and Arjona Agustin Perez and Skote Martin},
}

@article{TSOUTSANIS2023127544,
title = {A relaxed a posteriori {MOOD} algorithm for multicomponent compressible flows using high-order finite-volume methods on unstructured meshes},
journal = {Applied Mathematics and Computation},
volume = {437},
pages = {127544},
year = {2023},
issn = {0096-3003},
doi = {https://doi.org/10.1016/j.amc.2022.127544},
author = {Panagiotis Tsoutsanis and Machavolu Sai Santosh {Pavan Kumar} and Pericles S. Farmakis},
}

@ARTICLE{Tsoutsanis_climate,
author={Tsoutsanis, P. and Drikakis, D.},
title={A high-order finite-volume method for atmospheric flows on unstructured grids},
journal={Journal of Coupled Systems and Multiscale Dynamics},
year={2016},
volume={4},
pages={170-186},
doi={10.1166/jcsmd.2016.1104}
}

@ARTICLE{Tsoutsanis_cwenos,
author={Tsoutsanis, P. and Dumbser, M.},
title={Arbitrary High Order Central Non-Oscillatory Schemes on Mixed-Element Unstructured Meshes},
journal={Computer and Fluids},
year={2021},
volume={225},
doi={10.1016/j.compfluid.2021.104961},
art_number={104961}
}

@InProceedings{Tsoutsanis_eccomas2016_lmc,
author={Tsoutsanis, P. and  Simmonds, N. and Gaylard, A.},
title={IMPLEMENTATION OF A LOW-{M}ach NUMBER MODIFICATION FOR HIGH-ORDER FINITE-VOLUME SCHEMES FOR ARBITRARY HYBRID UNSTRUCTURED MESHES },
Booktitle={ECCOMAS Congress 2016, Crete, Greece},
year={2016},
doi={10.7712/100016.2004.8545},
}

@InProceedings{Tsoutsanis_hl,
author={Antoniadis, A.F. and Tsoutsanis, P. and Drikakis, D. },
title={Numerical Accuracy in {R}{A}{N}{S} Computations of High-Lift Multi-element Airfoil and Aircraft Configurations},
Booktitle={53rd AIAA Aerospace Sciences Meeting},
year={2015},
volume={0317},
doi={10.2514/6.2015-0317}
}

@ARTICLE{Tsoutsanis_HOVE1,
author={Ponweiser, T. and Tsoutsanis, P.},
title={Optimising {U}{C}{N}{S}{3}{D}, a High-Order Finite-Volume {W}{E}{N}{O} scheme code for arbitrary unstructured meshes},
journal={PRACE-HOVE 1 report},
year={2017},
url={https://prace-ri.eu/wp-content/uploads/WP222.pdf},
}

@ARTICLE{Tsoutsanis_HOVE2,
author={Shamakina, A. and Tsoutsanis, P.},
title={Optimisation of the Higher-Order Finite-Volume Unstructured Code Enhancement for Compressible Turbulent Flows},
journal={PRACE-HOVE 2 report},
year={2019},
url={https://prace-ri.eu/wp-content/uploads/WP288.pdf},
}

@InProceedings{Tsoutsanis_hyper,
author={Antoniadis, A.F and Drikakis, D. and Kokkinakis, I. W. and Tsoutsanis, P. and Rana, Z.},
title={High-Order Methods for Hypersonic Shock Wave Turbulent Boundary Layer Interaction Flow},
Booktitle={20th AIAA International Space Planes and Hypersonic Systems and Technologies Conference},
year={2015},
volume={3524},
doi={10.2514/6.2015-3524},
}

@ARTICLE{Tsoutsanis_knl,
author={Tsoutsanis, P.},
title={KNL Performance Comparison {U}{C}{N}{S}3{D}},
journal={ARCHER performance report},
year={2017},
pages={157-170},
url={www.archer.ac.uk/community/benchmarks/archer-knl/KNLperfUCNS3D.pdf},
}

@ARTICLE{Tsoutsanis_prace,
author={Tsoutsanis, P. and Antoniadis, A.F. and Jenkins, K.W.},
title={{Improvement of the computational performance of a parallel unstructured WENO finite volume CFD code for Implicit Large Eddy Simulation}},
journal={Computers and Fluids},
year={2018},
volume={173},
pages={157-170},
doi={10.1016/j.compfluid.2018.03.012},
}

@article{Tsoutsanis_rans,
title = {Assessment of high-order finite volume methods on unstructured meshes for {RANS} solutions of aeronautical configurations},
journal = {Computers and Fluids},
volume = {146},
pages = {86-104},
year = {2017},
doi = {https://doi.org/10.1016/j.compfluid.2017.01.002},
author = {Antoniadis, A. F.  and  Tsoutsanis, P. and Drikakis, D.},
}

@ARTICLE{Tsoutsanis_stencil,
author={Tsoutsanis, P.},
title={Stencil selection algorithms for {WENO} schemes on unstructured meshes},
journal={Journal of Computational Physics: X},
year={2019},
volume={4},
doi={10.1016/j.jcpx.2019.100037},
art_number={100037}
}

@ARTICLE{Tsoutsanis_weno,
author={Tsoutsanis, P. and Titarev, V.A. and Drikakis, D. },
title={{WENO} schemes on arbitrary mixed-element unstructured meshes in three space dimensions},
journal={Journal of Computational Physics},
year={2011},
volume={230},
number={4},
pages={1585-1601},
}

@ARTICLE{Tsoutsanis_weno_viscous,
author={Tsoutsanis, P. and Antoniadis, A.F. and Drikakis, D.},
title={{WENO} schemes on arbitrary unstructured meshes for laminar, transitional and turbulent flows},
journal={Journal of Computational Physics},
year={2014},
volume={256},
pages={254-276},
}

@misc{ucns3d,
  title = {{U}{C}{N}{S}{3}{D} {C}{F}{D} code},
  howpublished = {\url{http://www.ucns3d.com}},
  note = {Accessed: 2022-05-05},
}

@techreport{beckett2024archer2,
  author      = {Beckett, George and Beech-Brandt, Josephine and Leach, Kieran and Payne, Z{\"o}e and Simpson, Alan and Smith, Lorna and Turner, Andy and Whiting, Anne},
  title       = {{ARCHER2 Service Description}},
  institution = {Zenodo},
  year        = {2024},
  month       = dec,
  doi         = {10.5281/zenodo.14507040},
  url         = {https://doi.org/10.5281/zenodo.14507040}
}

@misc{lumi_hardware_2026,
  author       = {{LUMI User Documentation}},
  title        = {{LUMI System Architecture and Hardware Overview}},
  year         = {2026},
  howpublished = {\url{https://docs.lumi-supercomputer.eu/hardware/}},
  note         = {Accessed: 2026-07-30}
}

@misc{jsc_jupiter_technical_2026,
  author       = {{J{\"u}lich Supercomputing Centre}},
  title        = {{JUPITER Technical Overview}},
  year         = {2026},
  howpublished = {\url{https://www.fz-juelich.de/en/jsc/jupiter/tech/}},
  note         = {Last updated 26 June 2026; accessed: 2026-07-30}
}

@misc{jupiter_user_docs_2026,
  author       = {{J{\"u}lich Supercomputing Centre}},
  title        = {{JUPITER User Documentation: Hardware Overview}},
  year         = {2026},
  howpublished = {\url{https://apps.fz-juelich.de/jsc/hps/jupiter/configuration.html}},
  note         = {Accessed: 2026-07-30}
}

@misc{hlrs_hunter_2026,
  author       = {{High-Performance Computing Center Stuttgart}},
  title        = {{HPE Cray EX4000 (Hunter)}},
  year         = {2026},
  howpublished = {\url{https://www.hlrs.de/solutions/systems/hunter}},
  note         = {Accessed: 2026-07-30}
}

\end{document}